\documentclass[10pt,conference,compsocconf]{IEEEtran}
\usepackage{amsmath,amsthm}
\usepackage{amssymb}
\usepackage{algorithmic,algorithm}
\usepackage{multirow}
\usepackage{threeparttable}
\usepackage{array}
\usepackage{graphicx,color}
\usepackage[table]{xcolor}
\usepackage{subfigure}
\usepackage{listings}
\usepackage{caption}
\usepackage{pifont}
\usepackage{url}
\usepackage{array}
\usepackage[bookmarks=false]{hyperref}
\usepackage[skins,listings]{tcolorbox}
\usepackage{amsfonts}
\usepackage{float}
\usepackage{framed}
\usepackage{soul}
\usepackage{xcolor}
\usepackage{tikz}
\usepackage{pgf}
\usepackage{fancyvrb}

\usepackage{enumitem}
\usepackage{mathtools,amsthm}
\usepackage{cuted}
\usepackage{stackengine}
\usepackage{booktabs}
\usepackage{diagbox}
\usetikzlibrary{positioning}
\usetikzlibrary{decorations.pathreplacing}
\newcommand{\mypara}[1]{\vspace{.5em}\noindent\textbf{{#1. }}}

\newcounter{insight}[section]
\newcounter{question}[section]
\newenvironment{question}[1][]
{\refstepcounter{question}\par
\textbf{\texttt{Q\thequestion.#1}} \rmfamily}

\newcounter{challenge}[section]
\newenvironment{challenge}[2][]
{\refstepcounter{challenge}\par\medskip
   \textit{Challenge~\thechallenge : #2} #1 \rmfamily}
   
\newtcolorbox{codebox}[1]{
colback=white,
colframe=black,
fonttitle=\sffamily\bfseries\normalsize,
enhanced,
boxsep=0pt,
left=0pt,
right=0pt,
top=2pt,
title=#1}

\definecolor{mygray}{gray}{0.2}

\definecolor{tablegray}{gray}{0.9}
\lstdefinelanguage
    [x32]{Assembler}     
    [x86masm]{Assembler} 
    {morekeywords={movl,addl,cmpl,CMPXCHG16B,JRCXZ,LODSQ,MOVSXD, %
                  POPFQ,PUSHFQ,SCASQ,STOSQ,IRETQ,RDTSCP,SWAPGS,CALLQ,LEAVEQ,RETQ, %
                  rax,rdx,rcx,rbx,rsi,rdi,rsp,rbp, %
                  r8,r8d,r8w,r8b,r9,r9d,r9w,r9b, %
                  r10,r10d,r10w,r10b,r11,r11d,r11w,r11b, %
                  r12,r12d,r12w,r12b,r13,r13d,r13w,r13b, %
                  r14,r14d,r14w,r14b,r15,r15d,r15w,r15b}} 

\newcommand\MemoryLayout[1]{
  \begin{tikzpicture}[scale=0.3]
     \draw[thick](0,0)--++(0,3)node[above]{$0$};
     \foreach \pt/\col/\lab [remember=\pt as \tp (initially 0)] in {#1} {
       \foreach \a in {\tp,...,\pt-1} {
          \draw[fill=\col](-\a,0) rectangle ++(-1,2);
       }
       \draw[thick](-\pt,0)--++(0,3)node[above]{$\pt$};
       \if\lab\relax\relax\else
         \draw[thick,decorate, decoration={brace,amplitude=4mm}]
            (-\tp,-0.2)--node[below=4mm]{\lab} (-\pt,-0.2);
       \fi
     }
  \end{tikzpicture}
}

\begin{document}
\title{
\vspace*{3mm}
{\huge Spotting Silent Buffer Overflows in Execution Trace through Graph Neural Network Assisted Data Flow Analysis}
}
\author{\IEEEauthorblockN{Zhilong Wang, Li Yu, Suhang Wang
and Peng Liu}
\IEEEauthorblockA{College of Information Sciences and Technology\\
The Pennsylvania State University, USA\\
zzw169@psu.edu, szw494@psu.edu, luy133@psu.edu, pliu@ist.psu.edu}
}
\vspace{-1em}
\maketitle

\begin{abstract}

A software vulnerability could be exploited without any visible symptoms. When no source code is available, although such {\em silent} program executions could cause very serious damage, the general problem of analyzing silent yet harmful executions is still an open problem. In this work, we propose a graph neural network (GNN) assisted data flow analysis method for spotting  silent buffer overflows in execution traces. The new method combines a novel graph structure (denoted {\tt DFG+}) beyond data-flow graphs, a tool to extract {\tt DFG+} from execution traces, and a modified Relational Graph Convolutional Network as the GNN model to be trained. The evaluation results show that a well-trained model can be used to analyze vulnerabilities in execution traces (of previously-unseen programs) without support of any source code.
Our model achieves 94.39\% accuracy on the test data, and successfully locates 29 out of 30 real-world silent buffer overflow vulnerabilities. Leveraging deep learning, the proposed method is, to our best knowledge, the first {\em general-purpose} 
analysis method for silent buffer overflows. It is also the first method to spot silent buffer overflows in 
global variables, stack variables, or heap variables without crossing the boundary of allocated chunks. 


\end{abstract}

\section{Introduction}
A fundamental challenge in cybersecurity is that 
vulnerabilities widely exist in all kinds of programs~\cite{szekeres2013sok} despite 
software engineers and security analysts have been spending lots of efforts to avoid and test them.
These vulnerabilities could be exploited by the attackers and expose a huge threat to individuals, organizations and governments~\cite{krsul1998software}. 
Although researchers have proposed a variety of techniques to automatically discover and analyze software vulnerabilities~\cite{King:1976,sutton2007fuzzing,Xu:2016,arora2010empirical}, 
almost all existing techniques rely on visible ``symptoms" (e.g., crashes, failing assertions, and 
errors found by integrity checkers).  
Most vulnerability discovery methods use such symptoms to distinguish (potentially) harmful program executions
in which a vulnerability is triggered 
and benign executions~\cite{zalewski2014american,yun:qsym,chen2018angora}. 

However, a software vulnerability could be exploited without any visible symptoms, 
  and the corresponding program executions are often called a ``silent" yet harmful execution.  
For example, some silent buffer overflows, silent Use-After-Free, and 
  silent information leak could happen given specific malicious inputs. 
All these silent yet harmful program executions, though not as frequently seen as 
   harmful as executions carrying visible symptoms,  
could still be leveraged by attackers to compromise 
the system~\cite{chen2005non} and cause serious damage (e.g., altering program variables,
leaking critical information). 


When the source code is available, silent yet harmful executions in principle can be identified and analyzed.
By leveraging semantic information obtained from source code, researchers have developed various tools to 
identify and analyze them~\cite{serebryany2012addresssanitizer,jones1997backwards,1467813}.
For example, Konstantin et al.~\cite{serebryany2012addresssanitizer} developed AddressSanitizer to 
detect memory errors and diagnose root causes (of silent yet harmful executions)  
through source code level instrumentation.

However, in many cases the commercial software and legacy code targeted by the attacker 
  has {\bf no source code} available (to organizations using the software), and  
  it is widely recognized in the research community that when source code is not available,  
  analyzing silent yet harmful executions is an extremely difficult problem. 
A fundamental challenge in solving this problem is lack of high-level, semantically rich 
   information about data structures in the executables~\cite{7546500}. 
Due to the fundamental challenge, the {\bf general problem} of  
  identifying and analyzing silent yet harmful executions is yet to be solved. 
In the literature, only a small portion of silent vulnerabilities can be identified.
For example, there are a spectrum of silent buffer overflow, but 
 only overflow across the boundary of allocated chunks in heap can be 
 detected by existing methods~\cite{fioraldi2020fuzzing,dinesh2020retrowrite}. 
These methods capture the length of dynamically allocated buffer by hooking the heap allocation functions. Then they check the integrity of buffer access by comparing the buffer length and 
offset of buffer access. 
So far, no effective method has been proposed to analyze silent buffer overflows in 
  global variables, stack variables, or heap variables without crossing the boundary of allocated chunks.
As stated by Dinesh et al.~\cite{dinesh2020retrowrite}, binary disassembly is insufficient to recover data 
    section layouts and semantic information lost during compilation. 
Not surprisingly, silent vulnerability analysis without source code is 
  still an {\bf open problem} and there lacks a general purpose analysis method for 
  even one main category of silent vulnerabilities.  


In this work, we seek to develop a {\bf general purpose} analysis method for silent buffer overflows, one most important  
  category of silent vulnerabilities.  
The proposed method is based on a key observation: 

{\bf Key Observation.} {\em In silent yet harmful program executions, 
the data flows towards the variables corrupted by silent buffer overflow 
and the memory space layout of some of the corrupted variables 
are inherently different from those of non-affected variables.}
It is worth noting that human analysts have to examine enough data flows and memory layout patterns, which is usually 
very time consuming, before they could leverage this key observation and identify the exact 
differences between corrupted and non-affected variables.

In light of this, we propose to leverage Graph Neural Network~\cite{wu2020comprehensive} to significantly 
reduce manual efforts. In fact, our method is close to 100\% automatic. 
{\bf Our insight} is that critical information about the difference 
between corrupted and non-affected variables could be 
represented by a novel graph structure. 
Then Graph Neural Network (GNN) could learn essential features 
(from graphs extracted from execution traces)    
through representation learning.  
Then the learned representations could enable the GNN model to 
``analyze" a given execution trace by classifying the nodes in the graph extracted from the trace.   
Finally, the nodes classified as ``vulnerable" may provide enough information 
for automatically locating the address of vulnerable instructions and  
vulnerable  buffers, respectively.  


Specifically, we design a novel graph data structure to hold important features obtained from program executions, including data flows, variables' spatial information, and some useful implicit information flows. 
During the model training phase, we utilize a dynamic analyzer based on Intel Pin~\cite{luk2005pin} to build the newly designed graph automatically from execution traces (of various programs)  and customize AddressSanitizer~\cite{serebryany2012addresssanitizer} to help assign labels to nodes in the graph.
A node with label ``vulnerable" corresponds to a corrupted variable 
  resulted from silent buffer overflow.
Using the labeled graph as training data, 
  we design and train a Bi-directional Propagation Relational Graph Convolutional Network (\texttt{BRGCN}) to perform node classification. 

After the \texttt{BRGCN} model is well trained and deployed, the model can be used to analyze vulnerabilities in execution traces (of previously-unseen programs) without support of any source code.
The experiments show that our model achieves 94.39\% accuracy on the balanced test data, and successfully locates 29 out of 30 real-world vulnerabilities which we obtain from a public vulnerability database~\cite{cvedata,edb}. 
The evaluation results show that graph neural network assisted
data flow analysis is an effective general-purpose method in 
spotting silent buffer overflows when source code in not 
available. 

In summary, 
we made the following contributions:
\begin{itemize}
  \item To the best of our knowledge, this work proposes the first 
  graph neural network assisted data flow analysis method for
  spotting silent buffer overflows in execution traces. It 
  can analyze a full spectrum of silent buffer overflows. 
  \item We designed a new type of graph data structure \texttt{DFG+} to represent programs' data flow, variables' spatial information, and implicit information flow, in an integrated manner.  
  \item We implemented a tool based on Intel Pin to automatically generate \texttt{DFG+} from program executions and customized AddressSanitizer to help assign ground truth labels for nodes in \texttt{DFG+}. 
  \item We modified the Relational Graph Convolutional Neural Network (RGCN)~\cite{schlichtkrull2017modeling} by introducing bi-directional relation types to make it more effective in program analysis. 
  \item We evaluated the effectiveness of the newly designed \texttt{DFG+} 
  and the newly designed \texttt{BRGCN} model, and compared with other baseline methods. {\em In our view, the proposed method 
  is neither a ``competitor" nor an extension of existing fuzzing 
  tools.} Without source code, existing fuzzing tools, though very efficient, simply {\bf cannot} identify silent buffer overflows in 
  global variables, stack variables, or heap variables without 
    crossing the boundary of allocated chunks. 
  Hence, comparing the proposed method and fuzzing tools 
    could result in ``comparing apples and oranges." 
\end{itemize}

\section{Background and Related Work}
\label{sec:back}
\subsection{Buffer Overflow}
\label{sec:buff}
For decades buffer overflow (BOF) remains as one of the main security threats plaguing the cyberspace, attributing to the prevalence of the buffer overflow bugs in commodity software and the fundamental difficulty to find and fix them. 
Conventionally, BOF vulnerability refers to a category of software vulnerabilities which could corrupt the adjacent memory region due to insufficient bound checking.
The buffer associated with the vulnerability is called vulnerable buffer. 
According to the location of vulerable buffer, 
BOF can be group into heap, stack, and global BOF.
When BOF happens, it can cause the program to crash by corrupting data/code pointers (e.g., return address and jump table), 
or change program state by altering some non-control-data~\cite{chen2005non}.
In the paper, we classify \emph{wild} BOF into three categories according to its symptoms.
A \emph{wild} BOF means the input is not manually crafted by analyst, e.g., in exploit generation.

\begin{enumerate}[wide=\parindent,leftmargin=0.5em,topsep=0pt,itemsep=0ex,partopsep=0ex,parsep=0ex,label=\arabic*)]
\item \textit{Visible Buffer Overflow}.
We call a BOF visible if it shows visible symptoms, such as program crashing, assertion failing, or displaying garbled string on the screen.
\item \textit{Silent Buffer Overflow}. Another kind of BOFs happens without any visible symptoms. 
For instance, some BOFs that only corrupt some \textit{dead} variables (please see the define of \textit{dead} variable in Section~\ref{sec:dfg}), 
or only corrupt some local variable in stack, will not crash.
\item \textit{Innocent Buffer Overflow}. A BOF has no effect on the program if it only overwrites a padding space, 
which is inserted by some compilers to the boundary of buffers due to data alignment specified by attributes of variable~\cite{varalign} or language features~\cite{compileralign}. 
\end{enumerate}

In general, visible BOFs are believed to be easily discovered and analyzed when they happen, attributing to their visible symptoms. 
If source code is available, silent BOF can also be identified through sanitizer~\cite{serebryany2012addresssanitizer} or bound-checking~\cite{jones1997backwards}.
However, it is extremely challenging to analyze silent BOF when source code is unavailable, because critical high level information such as length of buffer and type of variables, is lost in the binary during the compilation.
The literature review shows that all the existing works~\cite{fioraldi2020fuzzing,dinesh2020retrowrite,bruening2011practical,reed1991purify,parallel} can only identify silent heap BOFs which overflow cross the boundary of allocated memory chunks.
The details of these approaches will be discussed shortly in Section~\ref{sec:relate}. 
Therefore, the general problem of silent BOF analysis is still an open problem.


\definecolor{dkgreen}{rgb}{0,0.6,0}
\definecolor{gray}{rgb}{0.5,0.5,0.5}
\definecolor{mauve}{rgb}{0.58,0,0.82}
\definecolor{mygray}{gray}{0.9}
\lstset{
language=C,
frame=tb,
basicstyle={\footnotesize \bf \ttfamily}, 
tabsize=3,
breaklines=true,
showstringspaces=false,
numbers=left,
numbersep=-10pt,                     
escapeinside={(*}{*)},
xleftmargin=2pt,
numberstyle=\texttt \footnotesize,
stringstyle=\color{mauve},
keywordstyle= \color{black}, 
commentstyle=\color{dkgreen} \textit,
}
\begin{figure}[b] 
\vspace{-4.5ex}
\begin{lstlisting}[language=C++,
    directivestyle={\color{black}}
    emph={int,char,double,float,unsigned},
    emphstyle={\color{blue}},caption={A piece of code with silent buffer overflow. },label=code:sbof,captionpos=b]
    int age, i, total = 0, ages[0x20];
    for(i = 0; i <= 0x20; i++){ 
        age = receive();
        if(age == -1) break;
        ages[i] = age; //overflow when i=0x20
        total += ages[i];
    }
\end{lstlisting}
\vspace{-2.5ex}
\end{figure} 
It is worth noting that certain BOF vulnerabilities in executiables
could display either visible symptom or invisible symptom in different executions, given different inputs. 
In this case if the vulnerability is triggered by one or more visible executions, the resulted BOF would be a visible BOF, even though the vulnerability could also be exploited by some silent executions.
Based on this fact, some works in program testing try to find vulnerabilities ``shared by'' visible BOFs and silent BOFs by varying input lengths~\cite{zalewski2014american,chen2018angora}.
However, this is only feasible when the vulnerability is caused by the excess length of inputs.
And it obviously cannot solve the general problem of silent BOF analysis. 
Firstly, many BOF vulnerabilities {\bf only} have silent BOF (execution) instances.
An illustrative example~\footnote{Note that some compilers may change the layout of variables in stack. In order to make the example simple, we do not consider variable reorder here.} is shown in Code~\ref{code:sbof}. 
In the code, {\small \bf \ttfamily ages} is an integer array of length 20 and {\small \bf \ttfamily i} is the index in the for loop to access {\small \bf \ttfamily ages}.
We could see there will be a BOF if {\small \bf \ttfamily i} equals {\small \bf \ttfamily 0x20} but the program won't terminate at this point.
Secondly, the length of buffer access does not necessarily depend on the length of input. 
It can depend on the length of a portion of the input, or the value of one or several bytes in the input. 
Under these circumstances, it is almost impossible to know which segment crashes the program.





\subsection{Data Flow Graph}
Data flow graph was first introduced in the data flow machines to describe parallel computation~\cite{hurson2007dataflow}.
A data flow graph is a directed graph, $G(N, E)$, where nodes in $N$ represent instructions 
and edges in $E$ represent data dependencies among the nodes.
Data flow graph is widely used in compiler optimization, such as register allocation, instruction scheduling and dead code elimination~\cite{kennedy1979survey}.
Although there is no universally accepted definition, data flow analysis 
generally refers to the process of collecting and deriving information
about the way the variables are defined, used in the program~\cite{badlaney2006introduction}.
 
Data flow graph and analysis are also widely used in software security to analyze software defects, enforce secure policies, and so on.
Compared with control flow graph, data flow graph is more informative, which contain 
semantic information of programs.





\subsection{Graph Neural Networks}



In recent years, deep neural networks have shown increasingly noticeable success in security domains, such as security-oriented program analysis~\cite{shin2015recognizing,chua2017neural,guo2019deepvsa} and anomaly detection~\cite{Du:2017:deeLog,loganomaly:2019}, due to their remarkable representation learning capabilities~\cite{bengio2013representation}. Some representative genres of deep neural networks are convolutional neural network (CNN), recurrent neural network (RNN), and graph neural network (GNN). CNN is developed to capture information from grid data, whereas RNN is designed to capture sequential information. Given the nature of our proposed graph data structure \texttt{DFG+}, GNN is more compelling because of its great ability in representation learning on graphs.

Convolutional graph neural networks (ConvGNN), among other GNNs, adopts convolution operations on graphs to capture local and global structural patterns, through designing special convolution and readout functions~\cite{zhang2020deep}. 
Standard convolutions on images or text embeddings are not applicable to graphs because graphs have irregular structures, so that special convolutions have to be designed to work on graph data~\cite{wu2020comprehensive}. 
Depending on how convolution is performed, existing GNNs can be classified into two categorizes, i.e., spectral-based convGNN and spatial-based ConvGNN~\cite{balcilar2020bridging}.

\subsubsection{Spectral-based ConvGNN}
Spectral-based convolution is defined based on spectral graph theory~\cite{chung1997spectral}.
In this framework, a graph Laplacian is defined and signals on graphs are filtered using eigen-decomposition of graph Laplacian.
The graph convolution operators are introduced by defining the graph Fourier transform. 
However, despite the solid mathematical foundations, such approaches suffer from large computational burden, spatially non-localized issue and generalization problem.
Considering the computation complexity of spectral-based ConvGNN, we don't adopt it in our analysis of \texttt{DFG+}. 

\subsubsection{Spatial-based ConvGNN}
The main idea of spatial-based ConvGNN (massage passing GNN) is to generate a node $v$'s representation by aggregating its own features $\mathbf{x}_{v}$ and neighbors' features $\mathbf{x}_{u}$, where $u \in $ set of neighbors of $v$.
Generally, spatial ConvGNN can be defined as:
\begin{equation}\label{eq:gcn}
    H^{(l+1)}=\sigma\left(\sum_{s} C^{(s)} H^{(l)} W^{(l, s)}\right),   
\end{equation}
where $H^{(l)} \in \mathbb{R}^{n \times d^{l}}$ is the latent representation of the $n$ nodes in the $l$ -th layer and $d^{l}$ is number of features. 
$C^{(s)}$ is the $s$-th convolution kernel that defines how the node features are propagated to the neighborhood nodes, $W^{(l, s)} \in \mathbb{R}^{d^{l} \times d^{l+1}}$ is the trainable weight matrix that maps the $d^{l}$-dimensional features into $d^{l+1}$ dimension, $\sigma$ is the activation function such as ReLU~\cite{nair2010rectified} and tanh.
Equation~\ref{eq:gcn} covers a broad class of ConvGNN, and different designs of spatial ConvGNN are distinguished by the their convolution kernel $C^{(s)}$ and variability induced by $W^{(l, s)}$ in Equation~\ref{eq:gcn}. 

In each layer, message passing algorithm (neighborhood aggregation) defined in the Equation~\ref{eq:gcn} aggregates features from a node's local neighborhood.
Therefore, the node representation learned by the a $k$-layer ConvGNN 
include not only the features of itself, but also the features of its $k$-hop neighborhoods and the local graph structure~\cite{zhang2020deep}.
The node representation learning ability of ConvGNN has been intensively investigated by recent research~\cite{xu2018powerful}.
The great ability of ConvGNNs in modeling graph structured data have facilitated various domains such as social network mining~\cite{qiu2018deepinf,tan2019deep}, knowledge graphs~\cite{wang2019knowledge}, bioinformatics~\cite{fout2017protein} and recently code similarity comparison~\cite{nair2020funcgnn}. Thus, it has great potential to adopt ConvGNNs for representation learning on \texttt{DFG+} to detect vulnerabilities.

\subsection{Related Works} 
\label{sec:relate}

\begin{table*}[ht]
  \footnotesize
  \setlength{\textfloatsep}{-2pt}
    \caption{Comparison of the related works' effectiveness to detect buffer overflow.}
    \label{tab:comp}
  \makeatletter
  \newcommand{\widenhline}{%
      \noalign {\ifnum 0=`}\fi \hrule height 0.5pt
      \futurelet \reserved@a \@xhline
  }
  \newcolumntype{I}{!{\vrule width 1pt}}
  \makeatother
  \centering
  \small
  \begin{tabular}{cIp{3.7cm}<{\centering}|p{2.7cm}<{\centering}|p{2.7cm}<{\centering}|p{2.7cm}<{\centering}}  
  \hline  
  \hline  
  Defence Tools &  Require Source Code or Not & Silent Heap BOF & Silent Stack BOF & Silent Global BOF \\ \widenhline 
  \hline  
  \textsc{Boil} & No & Partial & No & No \\ 
  \hline  
  AddressSanitizer & Yes & Yes & Yes & Yes \\ 
  \hline  
  TaintCheck & No & No & No & No \\   
  \hline  
  Memcheck & No & Partial & No & No \\
  \hline  
  Fuzzing & No & No & No & No \\   
  \hline  
  Symbolic Execution & No & No & No & No  \\  
  \hline 
  The Proposed Method & No & Yes & Yes & Yes  \\  
  \hline 
  \hline   
  \end{tabular} 
  \vspace{-0.3cm}
\end{table*}

In this section, we will introduce some works that spot BOF on the source code and binary level, and discuss their limitations. 

\mypara{Source Code based Schemes}
The source code based schemes usually adopt source code analysis to collect semantic information (e.g., buffer length) and enforce their detection rules through source code instrumentation. 
Among several existing works~\cite{serebryany2012addresssanitizer,jones1997backwards,1467813}, we choose one representative scheme~\textendash~AddressSanitizer (ASAN)~\textendash~to introduce. 
Briefly, ASAN detects spatial bugs by reserving \texttt{Redzone} around heap, stack, and global objects, detects temporal bugs by quarantine for heap and stack objects to delay the reuse of dead objects. 
Technically, it leverages \textit{shadow memory} to mark whether an address in the program space belongs to \texttt{Redzone} or not, and checks legality of target addresses before instructions access variables in memory.
Despite its effectiveness to detect all kinds of BOFs in global, stack and heap, ASAN has two limitations: 
firslty, ASAN needs program's source code, thereby cannot detect bugs in legacy code and commercial software.
Secondly, it fails to detect non-linear buffer-overflow (an access that jumps over a \texttt{Redzone}).

\mypara{Static Approaches on Binary}
Rawat et al.~\cite{rawat2012finding} researched the detection of potential stack-based BOF vulnerability in binary code.
Different from traditional works that usually define vulnerability patterns at the syntactic level (e.g., function name), they considered more features of vulnerabilities on semantic level and defined the buffer overflow inducing loops (\texttt{BOIL}) to summary the semantic patterns of potential vulnerable loop.
Based on the proposed patterns, they developed a prototype to identify potential vulnerable loops from executables.
The advantage of their approach is that it does not need to execute the program and can achieve high code coverage. However, as pointed out in their paper, their scheme can only deal with a special case of BOF. We think it is due to the challenges to summarize all patterns based on human efforts. 
In addition, the reported positive functions in this scheme can only be viewed as potential vulnerable functions and needs further verification, because no concrete input is available to verify the reported BOF.


\mypara{Dynamic Approaches on Binary}
Dynamic approaches analyze BOF vulnerabilities by finding vulnerable executions.
In the binary level, taint analysis (also known as data flow tracking) is a popular method to debug vulnerabilities. 
TaintCheck~\cite{newsome2005dynamic}, proposed by James Newsome and Dawn Song, which locates BOF based on one simple assumption~\textendash~in normal data flow, pre-defined taint source, such as user inputs, environment variables, and network data will not propagate to pointers.
Therefore, their approach can not detect the silent BOF which does not violate their assumption.

Memcheck~\cite{seward2005using} is a well-known vulnerability analysis tool that is implemented based on dynamic binary instrumentation framework Valgrind~\cite{nethercote2007valgrind}.
Technically speaking, it obtains the address and size of buffers in heap by hooking function calls to heap allocation functions and  
parsing their parameters and return values.
By comparing the offset of buffer accesses and the length of allocated buffer in heap, they detect heap BOF which writes out of allocated heap chunk.
As mentioned in~\cite{serebryany2012addresssanitizer}, Memcheck and some other tools (Dr. Memory~\cite{bruening2011practical}, Purify~\cite{reed1991purify} and Intel Parallel Inspector~\cite{parallel}) that adopt similar approaches are not capable to find out-of-bounds bugs in the stack (other than beyond the top of the stack frame), global and in heap if the overwrite does not across the boundary of allocated chunk.

Fuzzing~\cite{zalewski2014american}, symbolic execution~\cite{stephens2016driller}, gradient descent~\cite{chen2018angora} and hybrid approaches~\cite{yun:qsym} are widely adopted path exploration method to automatically find vulnerable paths in software.
These schemes try to generate input that achieve high coverage, and find input that can trigger bugs.
However, they select positive inputs based on whether it can crash the program.
In such a case, the silent BOFs are ignored.
Recent research works~\cite{fioraldi2020fuzzing,dinesh2020retrowrite} trying to detect silent vulnerabilities during Fuzzing suffer from same limitations with the works discussed in last paragraph.

Table~\ref{tab:comp} summary the limitations of related works we mentioned above. 
In conclusion, the general problem of silent BOF analysis is still an open problem, and it remains a fundamental challenge to cope with the {\bf full spectrum} of \textit{silent BOF} at the binary level.



\vspace*{-2mm} 
\section{Problem Formulation and Challenges}
\label{sec:problem}
In this section, we will formalize the silent BOF analysis problem and present the challenges to solve it.

\subsection{The Problem and Research Goal}
\label{sec:prob}
In this example shown in Code~\ref{code:sbof}, an integer array {\small \bf \ttfamily ages} with a fixed length is defined and allocated on the stack.
The loop copies one excess integer to the array and the nearest variable {\small \bf \ttfamily total} in the heap address will be overflowed, without disturbing the normal execution of the program.
We name the out-of-bound access (writing/reading) during execution as
\textit{invalid operations} or \textit{invalid access}, define the instruction address of the \textit{invalid operation} as the \textit{overflow point},
name execution with \textit{invalid operation} as \textit{BOF instance}, and a collection of runtime information as \textit{execution trace}. 

With the above notations, our research goal is to locate the \textit{overflow point} of silent BOF in an executable by analyzing its execution trace.
To be more specific, we want to:
\begin{enumerate}[wide=\parindent,leftmargin=0.5em,topsep=0pt,itemsep=0ex,partopsep=0ex,parsep=0ex,label=\arabic*)]
\item distinguish \textit{BOF instances} from normal executions.
\item locate the \textit{invalid operations} in \textit{execution trace} and \textit{overflow point} in an executable.
\item pinpoint each of them separately if there are more than one \textit{overflow points} in one \textit{execution trace}.
\end{enumerate}

\subsection{Why is This Problem Hard?}
\label{sec:chal}
\begin{figure}[b] 
\vspace{-2.5ex}
\lstset{
language=[x32]Assembler,
stringstyle=\color{black},
keywordstyle=\color{black},
}
\begin{lstlisting}[caption={The assemble code compiled from Code~\ref{code:sbof}.},label=code:asm,captionpos=b]
    sub    $0x94,%esp
    movl   $0x0,-0x10(%ebp)
    movl   $0x0,-0x14(%ebp)
    jmp    target2
    target1: call   <receive>
    mov    %eax,-0xc(%ebp)
    cmpl   $0xffffffff,-0xc(%ebp)
    je     target3
    mov    -0x10(%ebp),%ebx
    mov    -0x94(%ebp,%ebx,4),%eax
    add    %eax,-0x14(%ebp)
    addl   $0x1,-0x10(%ebp)
    target2: cmpl   $0x20,-0x10(%ebp)
    jle    target1
    target3:
\end{lstlisting}
\vspace{-2.5ex}
\end{figure}
Due to the unavailability of type information in binary, it is not possible to identify \textit{invalid operations} by comparing the offset of buffer reading/writing with length of target buffer.
As shown in Code~\ref{code:asm}, which is the assembly code generated from Code~\ref{code:sbof}, the instruction at line~1 allocates memory for variable {\small \bf \ttfamily age}, {\small \bf \ttfamily i}, {\small \bf \ttfamily total} and array {\small \bf \ttfamily ages},
and the instructions at line~2 and line~3 initialize two variables with {\small \bf \ttfamily 0x0}.
Number of allocated variables, the type and length of each variable are unrevealed in variable allocation and initialization instructions. 
Therefore, these lost information need to be inferred from the execution trace. For instance, we could develop heuristic rules to infer the boundary of buffers by 
distinguishing the data flow patterns to access buffers and to access their adjacent variables.

However, the inferences usually need a lot of domain knowledge associated with instruction set, conventions of compiler, and features of programming language. 
Generally, it is unclear which patterns the silent BOFs follow and which features exist in execution trace that could be used for analysis. In fact, designing heuristic rules is very time consuming and complicated~\cite{rawat2012finding} because it requires observing both enough silent BOF traces and normal execution traces.




\vspace*{-2mm} 
\section{Approach Overview}
\label{sec:over}

The challenges faced by traditional methods motivate us to solve the problem through deep neural networks.
Although the complexity of patterns to identify silent BOF is a daunting challenge for human analysts to develop heuristic rules, it may not be a challenge for a deep learning algorithm given enough training data.
Accordingly, we propose to spot silent BOF based on graph neural network assisted data flow analysis. 
In this section, we first provide several insights based on our domain knowledge, which motivate us to choose our technical approach, 
and these insights will be verified through several experiments. 
Then, we provide an overview of our proposed approach, and point out the challenges we must address. 

\begin{figure*}[!htbp]
    \setlength{\belowcaptionskip}{-13pt}
    \centering
    \includegraphics[width=1\textwidth]{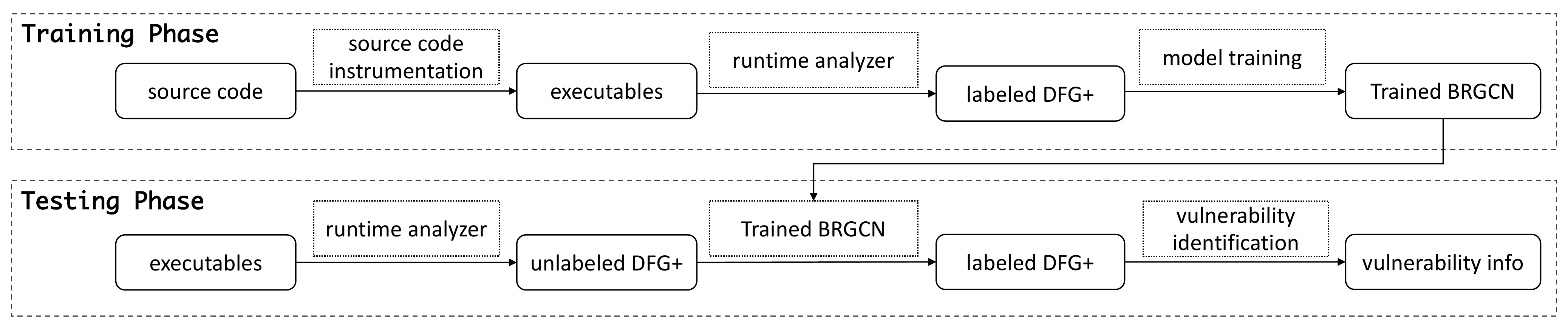}
    \caption{Approach overview.}
    \label{fig:ove}
 \end{figure*}
\subsection{Insights}

\subsubsection{The Essential Information Need to be Captured for BOF Analysis}
\label{sec:stru} 
Through dynamic binary instrumentation, lots of information can be collected along with program's execution, such as control flow, data flow, accessed memory, values of operands for each instruction, and executed instruction sequence.
However, not all these information are useful to identify silent BOF.
If the unnecessary information get included in training data, it will introduce noise and reduce the accuracy of the model. 
Hence, two questions are raised and the answers to them are associated with the domain knowledge related to buffer overflow and dynamic program analysis:

\begin{question}
\label{que:data}
What information should be selected?
\end{question}
\begin{question}
How to design the data structure to hold the data?
\end{question}

Firstly, as discussed in the Section~\ref{sec:buff}, most silent BOFs do not violate a valid control flow (by corrupted code pointers), but they always violate a valid data flow.
Therefore, the data flow is meaningful to be integrated to our training data. A data flow graph (DFG) is the most popular way to represent program's data flow.
Secondly, the spatial information (variable layout) is useful to diagnose BOF, because BOF is an \textit{spatial error}~\cite{szekeres2013sok}. We will discuss the challenge to represent the spatial information later on.

Thirdly, some other (implicit) information flow, such as information flow from a data pointer to its pointed variable, and information flow from condition variables to branch target, could be useful to infer whether variables are pointers or loop control variables.
These information could be of great importance as many BOFs happen due to unsafe the pointer dereference in a loop.

Fourthly, besides the information flow, information itself, i.e., the values of variables, could also be useful. 
In fact, the value of certain variables like loop control variables could be very useful to analyze BOF. However, 1) there is no deterministic relationship between value of loop control 
variables and BOFs, and 
2) the values of variables can be very noisy and hard to 
interpret on the binary level.  
Therefore, we decide not to include variable values in our data. Instead we propose to incorporate some attributes of variables such as whether a variable is immediate or is copied from the user input.



Based on the above insights, we leverage a novel graph data structure to capture the essential information. Since the graph we build is based on the program's runtime data flow graph, together with some spatial information, we call it Data Flow Graph Plus (\texttt{DFG+}). 

\subsubsection{Model Selection} 
\leavevmode

\begin{question}
Why is this a node classification problem? 
\end{question}
With the graph structure and its nodes and edges, graph analysis tasks can be grouped into three categories~\cite{wu2020comprehensive}: graph classification, node classification, and link prediction.

The graph classification aims to classify graphs into different types.
When it is applied to our \texttt{DFG+}, the problem becomes classifying whether a program execution contains silent BOF or not.
Since the goal of our work is not only to identify vulnerable execution, but also to locate the \textit{invalid operations} inside the execution, we cannot follow the graph classification task.
Link prediction is the problem of inferring missing relationships between entities (nodes), which also does not fit our need.
Node classification, on the other hand, aims to classify nodes into different categories.
If adopted, it could distinguish vulnerable nodes and benign nodes in graph, and the vulnerable point in execution trace can be located by mapping nodes from the graph to program trace.
Hence, the proposed research goals can be achieved by solving a node classification problem.
\leavevmode
\begin{question}
Why is graph neural network a promising approach?
\end{question}
Firstly, the graph neural network can learn node features and graph structure, which is exactly how the \texttt{DFG+} encodes data flow information.
Secondly, deep learn has shown very promising result in some reverse engineering works, such as~\cite{shin2015recognizing,chua2017neural}. 
In these works, it has shown superior performance compared with traditional methods, which indicates that it has great learning ability. 
Compared with other machine learning algorithms, the deep neural network has following 2 superiority:
As stated in~\cite{shin2015recognizing}, there are some attractive features of neural network, 
``first, neural networks can learn directly from the original representation with minimal feature engineering" and 
``second, neural networks can learn end-to-end, where each of its constituent stages are trained simultaneously in order to best solve the end goal".

\begin{question}
Why do we choose relational graph neural network?
\end{question}

Given the \texttt{DFG+} with multiple types of edges as the training data, it is natural to adopt the relational graph neural network. RGCN~\cite{schlichtkrull2017modeling} was originally proposed to represent knowledge bases with entities and triples as directed labeled multi-graphs. The entities are treated as nodes and the triples of the form (subject, predicate, object) are encoded by labeled edges, which is similar to the data structure in \texttt{DFG+}. There are other models capable of modeling graph structured data, that we think not suitable in our case. Graph recurrent neural network (GRNN)~\cite{seo2018structured} works on dynamic graphs where graphs are evolving over time~\cite{hajiramezanali2019variational}. The GraphSAGE network~\cite{hamilton2017inductive} does not consider different types of edges in node classification. The Heterogeneous graph neural network~\cite{zhang2019heterogeneous} aggregates heterogeneous attributes or contents associated with nodes, which is overly complicated for \texttt{DFG+}, where nodes contain relatively easy-to-encode attributes. Finally, label propagation~\cite{karasuyama2013multiple} relies on a nearest neighbor graph to generate pseudo-labels for nodes and is often used in semi-supervised learning~\cite{iscen2019label}.




\subsection{Overview}
\figurename{\ref{fig:ove}} provides an overview of our proposed method, which consists of two major phases.
In the training phase, we develop a runtime analyzer based on Intel Pin to trace program runtime information and organize it into a graph structure (i.e., \texttt{DFG+}). 
In the training samples, locations of \textit{invalid operations} in silent BOF execution are obtained through source code instrumentation. The \textit{invalid operations} are reflected as vulnerable labels for nodes in graph. We then train a \texttt{BRGCN} model on the labeled \texttt{DFG+} data, for testing in the following phase.
In the testing phase, the trained \texttt{BRGCN} model is used to predict silent BOF in programs with binary only.
The analyzer will trace program runtime information and construct \texttt{DFG+} without node label.
It also generates maps that mapping each node from \texttt{DFG+} to instructions in program and in execution trace. After labels are predicted for each node in \texttt{DFG+}, the mapping can help us to locate the \textit{invalid operation} in silent BOF.

\subsection{Challenges}
\vspace*{-2mm}

\begin{challenge}{How to apply the model trained on multiple programs/graphs to a previously-unseen program?}
\label{chall:multi} 
Due to the specificity of each program, selecting more semantic information from program execution inevitably introduce knowledge related to particular program logic into our dataset, which may not hold in other programs. 
These knowledge, if learned by the model, will hurt model's generalization ability.
As a result, some previous works~\cite{rajpal2017not,she2019neuzz} on neural network assisted Fuzzing can only let a model be trained and tested on the same program.
We will discuss how to cope this challenge shortly when presenting the design of \texttt{DFG+} and graph neural network in Section~\ref{sec:dfg} and Section~\ref{sec:model}, respectively.
\end{challenge}


\vspace*{-2mm}
\begin{challenge}{How to generate labels for \texttt{DFG+}?}  Generally, training a high quality model needs a fair amount of training samples with ground truth.
We do not want to manually label the nodes in \texttt{DFG+}, which requires lots of human efforts. Hence, we need to develop a tool to label the data samples automatically.
The detail of how data are labeled is presented in Section~\ref{sec:analyzer}.
\end{challenge}

\vspace*{-2mm}
\begin{challenge}{How to represent spatial information in a deep learning-friendly manner?}  Adding variable address to the training data could be the simplest way to include the spatial information. However, it is hard for the deep learning model to learn the variable layout from the variable addresses. We will discuss how we represent the spatial information so that the deep learning model can quickly capture it in Section~\ref{sec:dfg}.
\end{challenge}

\vspace*{-2mm}
\begin{challenge}{How to cope with extremely unbalanced dataset?} 
Each \texttt{DFG+} generated from program execution has more than 200,000 nodes on average, but only a few of them are vulnerability nodes, which means the dataset is extremely unbalanced.
\end{challenge}




\vspace*{-2mm} 
\section{Design and Implementation}
\label{sec:design}
In this section, we will firstly introduce the design of \texttt{DFG+}, technique details of compiler plugin and run-time analyzer, and how they work together to generate labeled \texttt{DFG+}.
Then, we present the \texttt{BRGCN} and how it helps to spot silent BOFs.

\subsection{DFG+}
\label{sec:dfg}

\subsubsection{Spatial Information}
The trained model should be able to capture the {\it general information} and ignore {\it program-specific information}, so that the model trained on one set of DFG+ can be applied to the other set. 
The {\it general information} is the knowledge shared among programs, such as the knowledge to determine whether two variables are adjacent to each other. 
The {\it program-specific information} is the knowledge that only comes with a specific program, for example an integer variable is located at \texttt{0x8048000}.

To encode spatial information, there are two potential methods: to integrate address of each variable into variable attributes in data flow graph or to use relations to reflect the adjacency relationships of variables.
We did not choose the first method due to two observations:
1) first, the spatial information, such as adjacency of two variable, are specific relations between variables, rather than entities or attributes, which should not be encoded as node features.
2) second, the value of variable address is always associated with a concrete execution and will change if a program is compiled with different compilers or options, or run in different system environment (e.g., different heap allocation), or even at different executions (e.g, different loading address due to ASLR~\cite{snow2013just}).
Integrating address into data flow graph will introduce {\it program-specific information} which is not helpful for the model.
Therefore, we instead use relations (edges in a \texttt{DFG+}) to indicate if two variables are adjacent to each other.
In this way, we can represent the spatial information in a deep learning-friendly manner.


\subsubsection{Basic Design}
Using the terminology from data flow analysis~\cite{khedker2017data}, a \textit{live} variable is \textit{defined} when an instruction writes value to the variable, and a \textit{live} variable is \textit{used} when an instruction reads the value of the variable.
A variable is \textit{live} at a program point $p$ if current value of this variable will be \textit{used} in future.
A \textit{live} variable $v$ is \textit{dead} at program point $q$ if after program point $q$ the value of $v$ is \textit{redefined} before it is \textit{used} or not will not be \textit{used} anymore.

\mypara{Nodes of Graph} A node in \texttt{DFG+} represents a \textit{live} variable.
Therefore, multiple nodes will created for a variable if the variable is \textit{defined} and \textit{redefined} along with program execution. 
Note that the ``variable" in our context not only refers to variables defined in source code, it can be operands of any instructions (e.g., the return address on stack, register and immediate value).
According to the attributes of the variable that a node corresponding to, we group nodes in graph into 4 types:
\begin{enumerate}[wide=\parindent,leftmargin=0em,topsep=0pt,itemsep=0ex,partopsep=0ex,parsep=0ex,label=\alph*)]
    \item[$-$] \texttt{Memory Node (m-node)} denotes a \textit{live} variable stored in memory.
    \item[$-$] \texttt{Register Node (r-node)} denotes a \textit{live} variable stored in register.
    \item[$-$] \texttt{Immediate Node (i-node)} denotes an immediate operand.
    \item[$-$] \texttt{External Node (e-node)} denotes a variable \textit{defined} by a system call. The \texttt{e-node} is a special type of nodes for variables associated to external data (e.g., user inputs and environment variables, and so on). The input data usually contain dangerous variables, which could result in BOF.
\end{enumerate}

\mypara{Edges of Graph} We define 5 classes of directed edges in \texttt{DFG+} to reflect program's direct or implicit information flow and spatial information. Note that each ``variable" that appears in following list is corresponding to a node in graph.
\begin{enumerate}[wide=\parindent,leftmargin=0em,topsep=0pt,itemsep=0ex,partopsep=0ex,parsep=0ex,label=\alph*)]
    \item[$-$] \texttt{Data Flow Edge (d-edge)} denotes a direct information flow from a source variable to a target variable. There exists a direct information if the value of the source variable is used to calculate the value of the target variable.

    \item[$-$] \texttt{Adjacency Edge (a-edge)} denotes that two variables are adjacent to each other. The direction of \texttt{a-edge} denotes relative high or low of two variable addresses.
   
    \item[$-$] \texttt{Index Edge (i-edge)} denotes an implicit information flow (implicit data flow). 
    The information flow is implicit if a pointer or offset $a$ is used to address a variable $b$ to be read or written. 

    \item[$-$] \texttt{Redefine Edge (r-edge)} denotes that a \textit{live} variable is covered by another \textit{live} variable.
    The \texttt{r-edge} not only indicates that these two \textit{live} variables are in the same address, but also implicates the order of data flow for this variable.

    \item[$-$] \texttt{Comparison Edge (c-edge)} denotes another kind of implicit information flow, which happens when a \textit{live} variable be compared with another \textit{live} variable. 
    The values of their operands will affect the value in \texttt{eflags} register, which then affect target of a conditional branch. 

\end{enumerate}

\begin{figure}[t]
    \centering
    \includegraphics[width=0.5\textwidth]{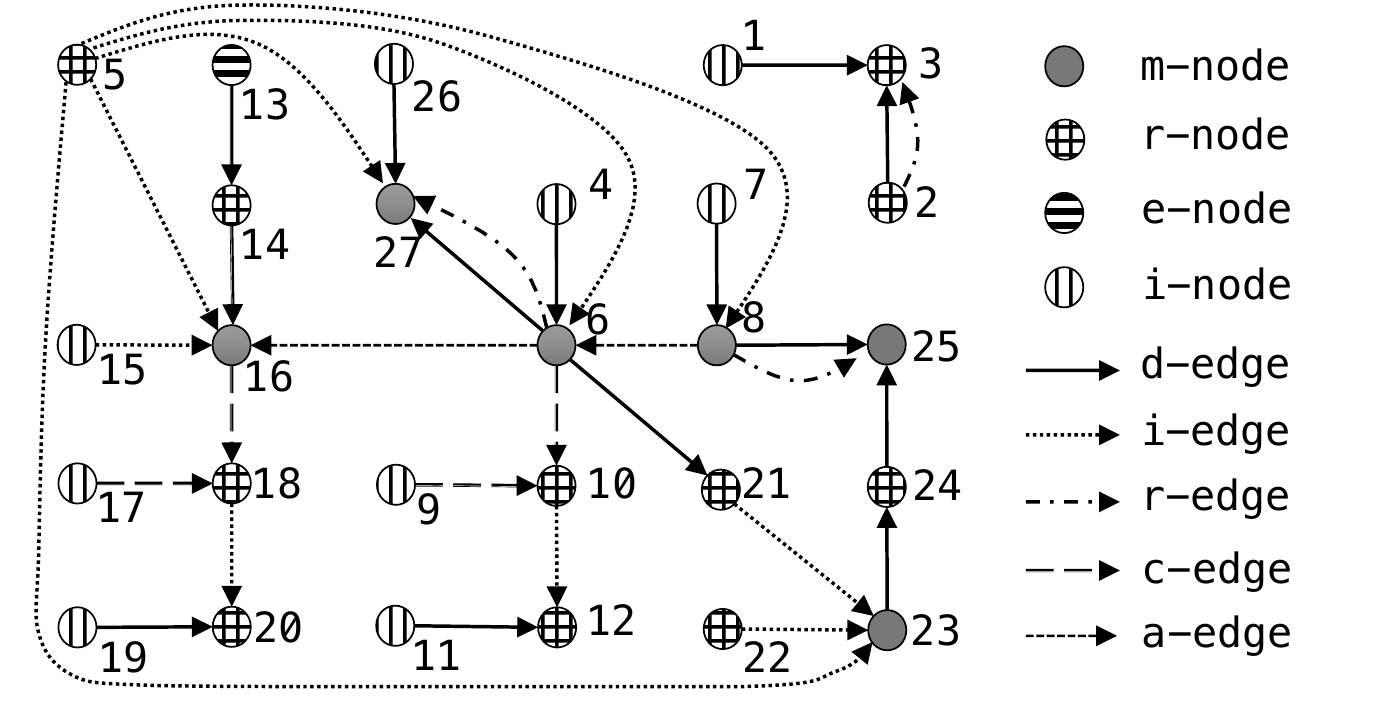}
    \caption{A DFG+ generated from execution of a piece of code.}
    \label{fig:dfgplus}
    \vspace*{-4mm}
\end{figure}

\figurename{~\ref{fig:dfgplus}} shows a \texttt{DFG+} generated from the execution of a piece of code in Code~\ref{code:asm}.
The executed instruction sequence is {\small \texttt{[$1,2,3,4,13,14,5,6,7,8,9,10,11,12$]}}.
Let's take the several nodes and edges generated from {\small \bf \ttfamily sub \$0x94, \%esp}, as an example to demonstrate how the graph is generated.
Node $1$, $2$, and $3$ , represent immediate {\small \bf \ttfamily \$0x94}, source operand {\small \bf \ttfamily \%esp}, and the destination operand {\small \bf \ttfamily \%esp}, respectively.
There is a \texttt{d-edge} and \texttt{r-edge} between $2$ and $3$ because the old value in {\small \bf \ttfamily \%esp} was used to calculate new value for {\small \bf \ttfamily \%esp}, and \textit{live} variable in {\small \bf \ttfamily \%esp} is \textit{redefined}.
Besides, the \texttt{a-edge} between node $6$ and $8$ denotes that \textit{live} variables in {\small \bf \ttfamily -0x10(\%ebp)} and {\small \bf \ttfamily -0x14(\%ebp)} are adjacent to each other.
The \texttt{i-edge} from node $5$ to node $6$ denotes that {\small \bf \ttfamily -0x14(\%ebp)} is used to address variable in {\small \bf \ttfamily -0x10(\%ebp)}.
The \texttt{c-edge} from node $6$ and $9$ to $10$ denotes that that the comparison between \textit{live} variable in {\small \bf \ttfamily -0x10(\%ebp)} and immediate number {\small \bf \ttfamily \$0x20} determines the value in {\small \bf \ttfamily \%eflags}.


\mypara{Labels of Nodes}
There are two types of labels for graphs nodes: 
1) \texttt{vulnerable label} represents that the node is generated from an \textit{invalid operation} in silent BOF;
2) \texttt{benign label} represents that it is generated from normal operation.

\subsubsection{Reflection}
Through the noval design, variable type, information flow and adjacency relationships of variables gathered through program tracing are represented by node features and graph structure through different types of edges (relations).
From the graph, we can not only clearly see the different graph structure associated with different type of operations,
but also see the difficulty to compare the difference between graph structures for vulnerable and benign nodes manually through human efforts.
In Section~\ref{sec:model}, we will show how Graph Neural Network capture these features through representation learning.

Here, we talk about how the design of \texttt{DFG+} helps to overcome Challenge~\ref{chall:multi}.
The novel design of \texttt{DFG+} aims to encode \textit{general information} and eliminate \textit{program-specific information} in program 
so that our trained model on some programs is able to be applied to other programs.
Specifically, variable address, variable value, and opcodes which are tightly associated with a specific program, 
are not included in \texttt{DFG+}.
Instead, we select address agnostic and value agnostic features~\textendash~information flow, variable adjacency and general variable features~\textendash~from execution trace, 
and encode them as different types of edges and node features in \texttt{DFG+}. 
So that a model trained on training set can applied to predict labels on the testing \texttt{DFG+}.






\subsection{Compiler Plugin for Data Labeling}
We implement a tool to insert some code to binary through source code instrumentation, which can automatically distinguish vulnerable and benign operations in program execution. As discussed in the related works, ASAN can detect out-of-bound memory accesses (i.e., \textit{invalid operations}) in BOF execution.
Therefore, we leverage ASAN to detect the \textit{invalid vulnerable operations}, which helps the graph constructor (to be discussed in next subsection) to label the nodes.
However, ASAN has four features, which pose 4 problems in our scenario: 1) ASAN inserts extra instructions before memory allocation, access and destroy. 2) ASAN inserts \textit{Redzone} among variables, which change the  adjacency relationships of variables. 3) ASAN reports memory errors by outputing vulnerable information, then terminate the execution. 4) ASAN can only detect BOF on function level for function linked from external libraries. Specifically, ASAN hooks function call to library functions, and provides wrapper functions to check whether BOF happens by analyzing the parameters passed to these library functions.

The extra instructions inserted by ASAN will introduce irrelevant information flow and the inserted \texttt{Redzones} break some \texttt{a-edge}s in the constructed \texttt{DFG+}. 
Besides, if the execution terminates at the point of first \textit{invalid access}, the data flow afterwards will be missing, so we have to modify ASAN to make it report \textit{invalid operations} without terminating the program's execution.
In the following paragraphs, we will show how we solve these problems.

\subsubsection{How To Exclude Irrelevant Data Flow from Instructions Inserted by ASAN?}
\label{sec:exclude}
To deal with the first problem, we need to distinguish program's original instructions and the instructions inserted by ASAN.
To achieve this goal, we modify the compiler plugin from ASAN to insert a pair of instructions (i.e., \texttt{prefetcht1} and \texttt{prefetcht2}) at the beginning and end of each piece of code inserted by the ASAN.
The pair of instruction serves as indicators that can be easily distinguished and skipped when the runtime analyzer builds \texttt{DFG+} along with program execution.
We adopt \texttt{prefetch} instructions because \texttt{prefetch} has no side effect to program's runtime state and we can easily disable them.


\subsubsection{How to Restore the Relation of Variable Adjacency?}
\label{sec:restoration}
To handle the second problem, we leverage the shadow memory to restore the original relation of variable adjacency.
Specifically, shadow memory maintained by ASAN's runtime environment recorder the location of inserted \texttt{Redzone} in the address space of target program. 
The compiler plugin will save configuration 
of shadow memory and share it to the graph constructor. 
Through these configuration, the graph constructor can query the shadow memory and restore the original adjacency relationships of variables.
We will discuss the details of how the adjacency relationships are restored in Section~\ref{sec:memrestore}.

\subsubsection{How to Label Nodes Generated from Vulnerable Operations?}
\label{sec:labeling}
To solve this problem, we let the compiler plugin to emit \texttt{prefetcha} as indicator before each suspicious instruction
(that results to out-of-bound read/write). 
Since ASAN checks the validity of target address before each suspicious memory access, and \texttt{prefetcha} will only be executed when memory errors are detected before it really happens.
By this way, the runtime analysis routine be notified through \texttt{prefetcha} and assigns different labels to nodes, accordingly. 
Thus, we can achieve our goal without terminating the program execution and introducing any irrelevant data flow. 



\subsubsection{How to Identify Vulnerable Operations in Library Functions?}
To solve the last problem, we instrument the necessary libraries with customized compiler, 
then link the instrumented library functions to target program.
However, we observe that the most commonly used library in linux~\textendash~glibc~\textendash~ 
cannot be compiled by LLVM due to some unsupported features, and the llvm-libc is still in planing phase~\cite{llvmlibc}.
Alternatively, we only instrumented vulnerable functions in glibc, such as \texttt{scanf} and \texttt{strcpy}.
Then, in the runtime library (runtime-rt~\cite{runtimert}) of LLVM, 
we hook calls to these vulnerable functions and redirect execution to instrumented ones.
In such a case, the vulnerable node in the glibc can be labeled accurately.

\textsc{Caveat.} The customized compiler plugin is only used to help the runtime analyzer 
to assign labels to vulnerable nodes in built graph.
The runtimer analyzer will assume all other nodes in graphs as benign nodes.
Therefore, there is no need to instrument other functions without vulnerabilities in libraries.
However, the memory allocation and free functions, such as \texttt{malloc} and \texttt{free}, are special cases.
Even through no BOF happens in these functions, we still need to instrument these functions to update the shadow memory.

\begin{figure*}[!htbp]
    \setlength{\belowcaptionskip}{-13pt}
    \centering
    \includegraphics[width=1\textwidth]{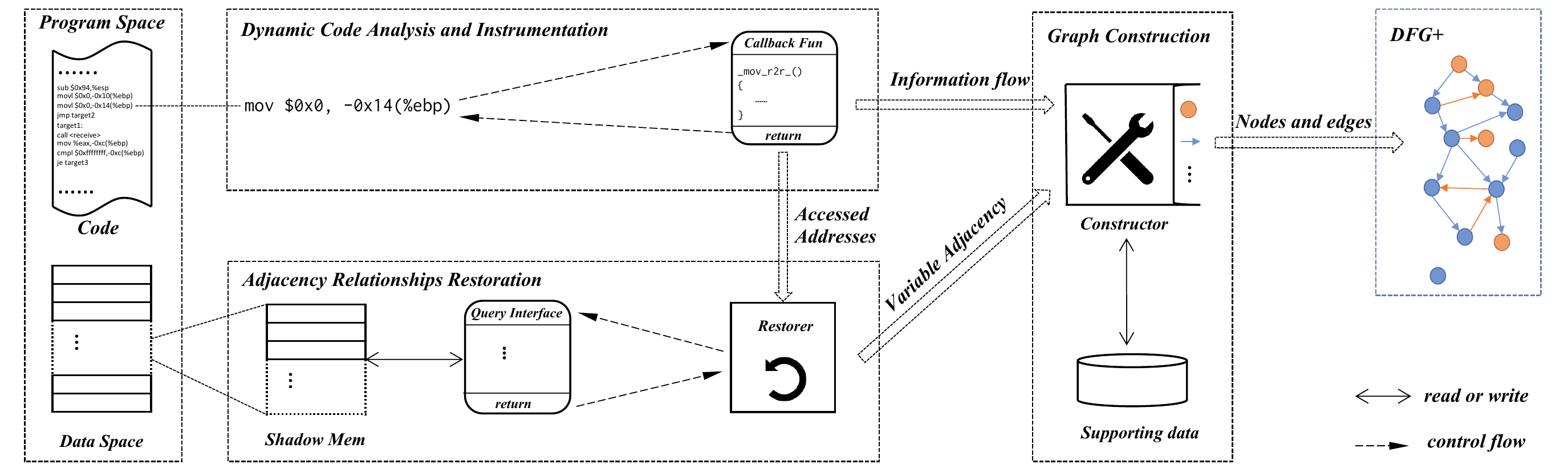}
    \caption{Workflow of the runtime analyzer to build \texttt{DFG+} with node labels.}
    \label{fig:analyzer}
 \end{figure*}
 
\subsection{DFG+ Construction based on Runtime Analyzer}
The runtime analyzer is implemented based on Intel Pin~\cite{luk2005pin}, 
which builds \texttt{DFG+} along with program's execution. 
Intel Pin provides comprehensive APIs for code inspection and instrumentation: the inspection APIs helps to analyze instructions in binary and the code instrumentation APIs help to instrument code according to the results of inspection.
The developed runtime analyzer consists of three components: 
dynamic code analysis and instrumentation, memory layout restoration, and graph construction.
\figurename{~\ref{fig:analyzer}} demonstrates the whole workflow.

\subsubsection{Dynamic Code Analysis and Instrumentation}
\label{sec:analyzer}
The dynamic code instrumentation consists of three phases: code inspection, code instrumentation and runtime analysis.
Before code instrumentation the analyzer firstly analyzes instructions and system calls.
Three types of callback functions will be registered according to the analysis results:

\begin{itemize}
    \item  \texttt{Instruction Callback.}
    The structure of information flow can be easily understood given some examples: the code analysis routine uses \texttt{1memory-to-register} and \texttt{2register-to-register} to define the structures of information flow in {\small \bf \ttfamily mov 0x8048000, \%eax}, {\small \bf \ttfamily sub \%eax, \%ebx}, respectively.
    Then, callback functions are registered to instructions according to the types and structures of information flow as demonstrated in \figurename{~\ref{fig:analyzer}}.

    \item \texttt{System Call Callback.}
    Some system calls copy some external data to program space, the variables in which should be recognized as \texttt{e-nodes}. Call back functions are registered to these system calls to label corresponding memory regions at runtime.
    
    \item \texttt{Control Callback.}
    Two callback functions should be registered to \texttt{prefetcht1} and \texttt{prefetcht2} to stop and resubmit the runtime tracing respectively, so that inserted pieces of code by ASAN can be skipped.
    A callback function should be registered to \texttt{prefetchta} to receive the signal about \textit{invalid operations}, and assign labels to corresponding vulnerable nodes accordingly. 
    
\end{itemize}

The compiler plugin based on LLVM instruments code on the intermediate representation (IR) during compilation.
During experiments, we observe that some instructions reside outside of \texttt{prefetcht1}-\texttt{prefetcht2} pair in IR level during source code instrumentation float to position which are enclosed by \texttt{prefetcht1}-\texttt{prefetcht2}, due to instruction reordering~\cite{heisch1999method}.
In such case, information flow resulting from the floated instructions will be lost if we simply stop the analysis process when execution enters code enclosed by \texttt{prefetcht1}-\texttt{prefetcht2}.

To solve the problem, we adopt static data flow analysis 
to identify the floated instructions insides the \texttt{prefetcht1}-\texttt{prefetcht2} pair
based on one heuristic rule: an instruction $i$ inside \texttt{prefetcht1}-\texttt{prefetcht2} 
pair is a floated instruction if there is a data dependency between an instructions $j$ which is after
\texttt{prefetcht2} and $i$.
Accordingly, we will not exclude information flow resulting from the floated instructions in runtime analyzer.
Then, along with program execution, the callback functions will capture the information flow and access memory addresses from executed instructions,
and send them to the graph constructor and  adjacency relationships restorer.

\subsubsection{Adjacency Relationships Restoration}
\label{sec:memrestore}
\begin{figure}[b]
\captionsetup{justification=centering}
\footnotesize
\setlength{\belowcaptionskip}{-4pt}
\centering 
    \begin{tikzpicture}[scale=0.3]
        \draw[fill=white!30](2,0) rectangle ++(-2,2) node[pos=.5]{...};
        \draw[fill=white!30](-16,0) rectangle ++(-2,2) node[pos=.5]{...};
        \draw[thick,white](2,0)--++(0,2);
        \draw[thick,white](-18,0)--++(0,2);

        \foreach \a in {0,...,3}{
            \draw[fill=blue!10](-\a,0) rectangle ++(-1,2);
        }
        \draw[thick,decorate, decoration={brace,amplitude=4mm}]
                (-0,-0.2)--node[below=4mm]{\footnotesize var2} (-4,-0.2);
        
        \draw[thick](-4,0)--++(0,3)node[above]{{\scriptsize $i$+12}};
        \foreach \a in {4,...,11}{
            \draw[fill=orange!20](-\a,0) rectangle ++(-1,2);
        }
        \draw[thick,decorate, decoration={brace,amplitude=4mm}]
                (-4,-0.2)--node[below=4mm]{\footnotesize buf1} (-12,-0.2);

        \draw[thick](-12,0)--++(0,3)node[above]{{\scriptsize $i$+4}};
        \foreach \a in {12,...,15}{
            \draw[fill=yellow!30](-\a,0) rectangle ++(-1,2);
        }
        \draw[thick,decorate, decoration={brace,amplitude=4mm}]
                (-12,-0.2)--node[below=4mm]{\footnotesize var1} (-16,-0.2);
        \draw[thick](-16,0)--++(0,3)node[above]{{\scriptsize $i$}};
    
        \draw [-,black,densely dotted] (-12.5, 2) to [out=90,in=90] (-11.5, 2);
        \draw [-,black,densely dotted] (-8.4, 2) to [out=90,in=90] (-7.5, 2);
        \draw [-,black,densely dotted] (-9.5, 2) to [out=90,in=90] (-8.6, 2);
        \node [below=10mm] at (-7,0){\small (a) Original memory layout (ML).} ;
\end{tikzpicture}
\begin{tikzpicture}[scale=0.24]
    \draw[fill=white!30](2,0) rectangle ++(-2,2) node[pos=.5]{...};
    \draw[fill=white!30](-32,0) rectangle ++(-2,2) node[pos=.5]{...};
    \draw[thick,white](2,0)--++(0,2);
    \draw[thick,white](-34,0)--++(0,2);
    \foreach \a in {0,...,3}{
        \draw[fill=red!30](-\a,0) rectangle ++(-1,2);
    }
    \draw[thick,decorate, decoration={brace,amplitude=4mm}]
            (-0,-0.2)--node[below=4mm]{\footnotesize red4} (-4,-0.2);
    \draw[thick](-4,0)--++(0,3)node[above]{{\scriptsize $j$+28}};
    \foreach \a in {4,...,7}{
        \draw[fill=blue!10](-\a,0) rectangle ++(-1,2);
    }
    \draw[thick,decorate, decoration={brace,amplitude=4mm}]
            (-4,-0.2)--node[below=4mm]{\footnotesize var2} (-8,-0.2);
    \draw[thick](-8,0)--++(0,3)node[above]{{\scriptsize $j$+24}};
    \foreach \a in {8,...,11}{
        \draw[fill=red!30](-\a,0) rectangle ++(-1,2);
    }
    \draw[thick,decorate, decoration={brace,amplitude=4mm}]
            (-8,-0.2)--node[below=4mm]{\footnotesize red3} (-12,-0.2);
    \draw [->,violet!80] (-10.5, 2) to [out=90,in=90] (-6.5, 2);
    \draw[thick](-12,0)--++(0,3)node[above]{{\scriptsize $j$+20}};
    \foreach \a in {12,...,19}{
        \draw[fill=orange!20](-\a,0) rectangle ++(-1,2);
    }
    \draw[thick,decorate, decoration={brace,amplitude=4mm}]
            (-12,-0.2)--node[below=4mm]{\footnotesize buf1} (-20,-0.2);
    \draw[thick](-20,0)--++(0,3)node[above]{{\scriptsize $j$+12}};
    \foreach \a in {20,...,23}{
        \draw[fill=red!30](-\a,0) rectangle ++(-1,2);
    }
    \draw[thick,decorate, decoration={brace,amplitude=4mm}]
            (-20,-0.2)--node[below=4mm]{\footnotesize red2} (-24,-0.2);
    \draw[thick](-24,0)--++(0,3)node[above]{{\scriptsize $j$+8}};
    \foreach \a in {24,...,27}{
        \draw[fill=yellow!30](-\a,0) rectangle ++(-1,2);
    }
    \draw[thick,decorate, decoration={brace,amplitude=4mm}]
            (-24,-0.2)--node[below=4mm]{\footnotesize var1} (-28,-0.2);
    
    \draw [-,black,densely dotted] (-24.5, 2) to [out=90,in=90] (-19.5, 2);
    \draw [-,black,densely dotted] (-16.4, 2) to [out=90,in=90] (-15.5, 2);
    \draw [-,black,densely dotted] (-17.5, 2) to [out=90,in=90] (-16.6, 2);
    \draw[thick](-28,0)--++(0,3)node[above]{{\scriptsize $j$+4}};
    \foreach \a in {28,...,31}{
        \draw[fill=red!30](-\a,0) rectangle ++(-1,2);
    }
    \draw[thick,decorate, decoration={brace,amplitude=4mm}]
            (-28,-0.2)--node[below=4mm]{\footnotesize red1} (-32,-0.2);
    \draw[thick](-32,0)--++(0,3)node[above]{{\scriptsize $j$}};

    \node [below=10mm] at (-15,0){\small (b) Memory layout (ML) after code instrumentation (w/ \texttt{Redzone}).} ;
    \end{tikzpicture}
\caption{Comparison between memory layouts with and without \texttt{Redzone}.}
\label{fig:layout}
\end{figure}
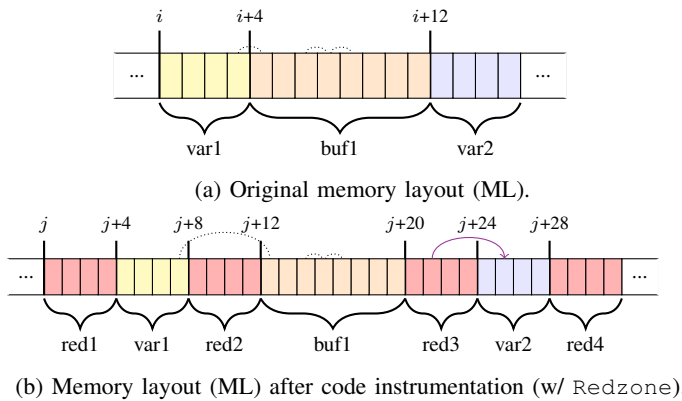

We observe that there are three kinds of changed adjacency relationships of variables, requiring different treatments respectively.
We will show these three cases based on \figurename{~\ref{fig:layout}}, which shows the layout of a memory fragment before and after instrumentation by ASAN.

Firstly, case~\ding{172} does not need any restoration. 
For byte(s) inside a buffer or variable, the inserted \texttt{Redzones}
do not affect its adjacency relationships, thereby needing no restoration. 

Secondly, case~\ding{173} needs restoration. 
For byte(s) on the boundary of a buffer or variable, 
its adjacency relationships get changed because of the inserted \texttt{Redzone}.
For example, the adjacent bytes of $i$+4 in ML is the byte $i$+3
and byte $i$+5.
However, in the ML w/ \texttt{Redzone}, the adjacent bytes for $j$+12 
is byte $j$+11 and $j$+13, and the byte $j$+11 is located in the \texttt{Redzone}.
To restore adjacency relationships for this kind of byte(s), 
we find the real adjacent bytes by skipping bytes in \texttt{Redzone}.
By skipping \texttt{Redzone2}, the real adjacent byte, i.e., $j$+7, of byte $j$+12 can be found.

Thirdly, case~\ding{174}, which happens in BOF, also needs restoration.
When out-of-bound access happens in a ML w/ \texttt{Redzone}, one or several byte(s) (e.g., $x$) in \texttt{Redzone} will be read/written.
If the \textit{invalid access} is mapped to the ML, the out-of-bound read/write will read/write a byte(s) near the vulnerable buffer in high address.
The following three steps can help to find the corresponding byte(s) in ML w/ \texttt{Redzone}:
\begin{enumerate}[wide=\parindent,leftmargin=0.5em,topsep=0pt,itemsep=0ex,partopsep=0ex,parsep=0ex,label=\arabic*)]
\item First, finding the boundary byte ($b$) of BOF, which is the byte near the first overflow byte(s) in low address.

\item Second, calculating the distance ($d$) between $b$ and $x$.

\item Third, finding byte(s) ($y$) by shifting $d$-1 byte from the $b$ to higher address, while skipping all bytes in \texttt{Redzone}.
\end{enumerate}

After mapping the byte(s) $x$ in \texttt{Redzone} to a byte(s) $y$ outside of \texttt{Redzone}, 
the adjacent bytes found through strategies adopted in case \ding{172}, \ding{173}, and \ding{174} for $y$ is the 
restored adjacent bytes for $x$. 
For example, through aforementioned strategies, 
we can map byte in $j$+21 to byte in $j$+25 and 
find its real adjacent bytes in $j$+24 and $j$+26.

\subsubsection{Graph Construction}
After the information flow are captured and filtered through callback functions at runtime, 
and the adjacency relationships are restored through aforementioned three strategies,
it is straightforward to construct \texttt{DFG+}.
We will not cover the details of graph construction.

\mypara{Supporting Data}
Some data, named as supporting data as shown in \figurename{~\ref{fig:analyzer}}, is important not only in the graph construction phase, but also in vulnerability identification phase.
For example, 1) $map$ that maps a node in \texttt{DFG+} to the address of its corresponding variable, 
2) $map$ maps a node to the instruction, which creates the node, and so on.
We will show how these information is used in Section~\ref{sec:iden}.



\textsc{Caveat.} After the model is trained, we no longer need source code to generate labels, as the model will predict them for us. The building the unlabeled graphs for binary-only programs in testing phase as shown in \figurename{~\ref{fig:ove}}, is much easier.
Since the analyzed programs are not to be instrumented, there is no need to exclude irrelevant instructions and restore the adjacency relationships of variables.




\subsection{Our Graph Neural Network}
\label{sec:model}
\texttt{DFG+} is a novel graph data structure to hold variable attributes, program information flow and variable layout. Generally, the vulnerable data flow in the execution context and the variable layout for some variables corrupted by silent  buffer overflow could be different from that of non-affected variables. In other word, the local graph centered at a vulnerable node would be slightly different from that of benign node. Thus, detecting vulnerability is equivalent to node classification by considering the local graph centered at each node. Thus, we need to design a model that's able to learn node representations that capture the local graph structure and neighborhood information, which facilitates the differentiation of vulnerable nodes from benign nodes. 
Thanks to the message passing mechanism, GNNs are  good at learning node representations by aggregating a node's neighborhood information. Thus, we adopt GNNs for \texttt{DFG+}. As \texttt{DFG+} has different types of nodes and edges,  we propose to adopt RGCN~\cite{schlichtkrull2017modeling} as our basic model because it is developed for representation learning in knowledge graph, which also have different types of nodes and edges.


Essentially, the multiple layer RGCN learns node representation for a node $v_x$ by aggregating features (attributes) of node $v_x$ and its neighbors through message passing. In particular, for different types of edges/nodes, it use different parameters during message passing, thus preserving the edge/node type information. 
\begin{figure*}[!htbp]
    \setlength{\belowcaptionskip}{-13pt}
    \centering
    \includegraphics[width=0.95\textwidth]{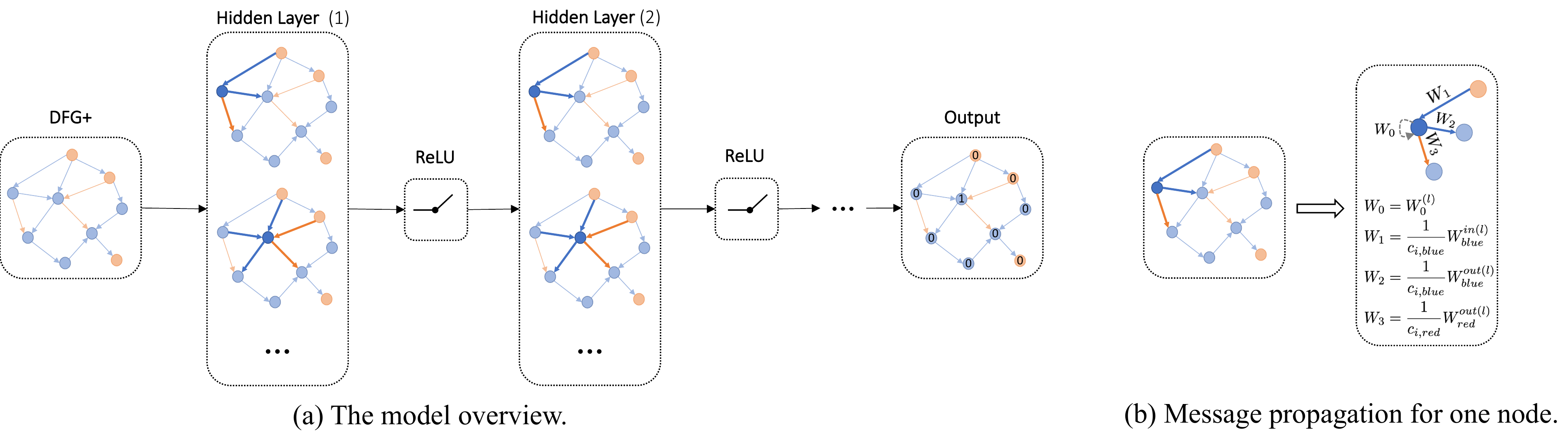}
    \caption{Bi-directional Relational Graph Convolutional Network.}
    \label{fig:birgcn}
\end{figure*}
The propagation rule of RGCN in the $l$-th layer for calculating the forward-pass update of a node $v_i$ is:
\begin{equation}\label{eq:rgcn}
    h_{i}^{(l+1)}=\sigma\left(\sum_{r \in \mathcal{R}} \sum_{j \in \mathcal{N}_{i}^{r}} \frac{1}{c_{i, r}} W_{r}^{(l)} h_{j}^{(l)}+W_{s}^{(l)} h_{i}^{(l)}\right),
\end{equation}
where $\mathcal{N}_{i}^{r}$ denotes the set of neighbor indices of node $i$ under relation $r \in \mathcal{R}$ and $\mathcal{R}$ is a set of relation (edge types). $c_{i, r}$ is a normalization constant that we set as the count of neighbor relation $r$ for node $i$.  
$W_{r}^{(l)}$ is relation-specific transformations matrix for relation $r$, which enables relation-specific message passing, thus preserving edge type relationship.
To ensure that the representation of a node at layer $l$ + 1 can also be informed by the corresponding representation at layer $l$, a single self-connection (i.e., $W_{s}^{(l)}$) term is added.
All messages passed along with incoming edges are aggregated through an element-wise activation function $\sigma(\cdot)$. 
$W_{r}^{(l)}$ and $W_{s}^{(l)}$ are the parameters to be learned.  By stacking $K$-layers of RGCN together, the representation of node $v_i$ could capture the $K$-hop local graph information centered at node $v_i$.

\mypara{Limitation of RGCN} 
Equation~\ref{eq:rgcn} is the basic design of RGCN, which has shown promising result in the early research~\cite{schlichtkrull2017modeling}.
For node classification in \texttt{DFG+}, however, the features of a node $x$'s outgoing nodes are not used in an appropriate way. If $\mathcal{N}_{i}^{r}$ is defined as the set of neighbor indices of node $v_i$ under relation $r$ through incoming edge, message can only pass along in these directions.
As a consequence, node representation learned by the network only aggregates features from incoming nodes and some important features from outgoing nodes are lost.
For example, if a global variable is \texttt{used} twice at runtime, 
its corresponding node $x$ in the \texttt{DFG+} will have two outgoing \texttt{d-edges} to node $y$ and $z$ respectively, i.e., \textcircled{$y$} $\leftarrow$ \textcircled{$x$} $\rightarrow$  \textcircled{$z$}.
In such case, the feature cannot be propagated from $y$ to $z$ or from $z$ to $y$ through $x$, which is undesired because node $y$ could be very useful to classify node $z$ and vice versa. For nodes without incoming links such as $x$ in the previous example, no information will propagate to them and thus we cannot learn good representations.

\mypara{RGCN with Bi-directional Propagation}
One straightforward solution to the above issues would be ignoring the direction of the edge, i.e., if the $\mathcal{N}_{i}^{r}$ is defined as the set of neighbor indices of node $i$ under the relation $r$ through either incoming edge or outgoing edge, messages will get processed with the same relation-specific transformations $W_{r}^{(l)}$.
However, this will ignore the difference of incoming and outgoing directions.

From the observations above, we extend the basic design to a bi-directional propagation for directed graphs.
Specifically, we adopt two set of parameter for each type of edge: 
\begin{enumerate}[wide=\parindent,leftmargin=0.5em,topsep=0pt,itemsep=0ex,partopsep=0ex,parsep=0ex,label=\arabic*)] 
\item $W_{r}^{in}$ is used to propagate messages along with the direction of directed edge;
\item $W_{r}^{out}$ is used to propagate messages against the direction of directed edge;
\end{enumerate}
We define the propagation rule as:
\begin{equation}
    \label{eq:rgcn2}
    \begin{split}
        h_{i}^{(l+1)} &= \sigma\left(\sum_{r \in \mathcal{R}} \left( \sum_{j \in \mathcal{IN}_{i}^{r}} \frac{1}{c_{i, r}} W_{r}^{in(l)} h_{j}^{(l)} + \right.\right.\\
         &  \left.\left. \sum_{k \in \mathcal{OUT}_{i}^{r}} \frac{1}{c_{i, r}} W_{r}^{out(l)} h_{k}^{(l)} \right) + W_{0}^{(l)} h_{i}^{(l)} \right) 
    \end{split}
\end{equation}
where $\mathcal{IN}_{i}^{r}$ and $\mathcal{OUT}_{i}^{r}$ denote the set of incoming neighbors and outgoing neighbors for node $i$ under the relation $r \in \mathcal{R}$, respectively.
The transformations $W^{(l)}$ is applied based on the type and direction of edge. 
By designing two sets of weights for both directions, we make sure that the information of node $y$ has a chance to propagate to node $z$ and vice versa (the example shown in the last subsection).
In our evaluation, we will quantitatively evaluate the model represented by Equation~\ref{eq:rgcn2} and compare its effectiveness with the basic design denoted by Equation~\ref{eq:rgcn}.

Moreover, we deprecate the common one-hot encoding of IDs for each node as adopted by~\cite{schlichtkrull2017modeling}.
Instead, we make the type of node as the node features, and expect the model to behave the same regardless of the node order.
We will evaluate its effectiveness in Section~\ref{sec:explain}.


\figurename{~\ref{fig:birgcn}(a)} shows the framework of \texttt{BRGCN}, which takes \texttt{DFG+} as input and predict the labels for each nodes.
\texttt{DFG+} consists of difference types of nodes and edges, which are marked with different colors in the figure.
The $W$ and $c$ in Equation~\ref{eq:rgcn2} are parameters of the model, which is learned during the training phase.
Initially, node features in \texttt{DFG+} are embedded and fed into the model as the input of the first layer. 
Then, layer $l$ computes the update feature (latent node representation $h^{(l+1)}$) for each node $v_i$ by aggregating features from its neighbors and itself. 
The output of the previous layer become the input to the next layer.
Finally, in the output layer, $softmax(\cdot)$ activation is applied to generate label probabilities.

\figurename{~\ref{fig:birgcn}(b)} illustrates message-passing when calculating update feature for a node $i$ in layer $l$.
Features from neighboring nodes are gathered and then transformed for each relation type individually, together with different transformation matrix $W$s for different types of edges.
For example, $W_{blue}^{out(l)}$ is the transformation matrix for outgoing blue edges.
The resulting representations are accumulated and normalized.
We choose \texttt{ReLU} as activation function in our model.

\texttt{DFG+} is a directed graph $G=(\mathcal{V}, \mathcal{F}, \mathcal{E}, \mathcal{R})$ with nodes (entities) $v_{i} \in \mathcal{V}$ have feature $f_{i} \in \mathcal{F}$, and edges with (relations) $\left(v_{i}, r, v_{j}\right) \in \mathcal{E}$, where $r \in \mathcal{R}$ is a relation type.
In each layer $l$, node features are updated through function defined in Equation~\ref{eq:rgcn2}. 
For each node, its old features ($h_i^{(l)}$) and its neighbors' old features ($h_j^{(l)}$) are passed along with the edges ($\left(v_{i}, r, v_{j}\right) \in \mathcal{E} \lor \left(v_{j}, r, v_{i}\right) \in \mathcal{E}$), and then aggregated through a normalized sum ($\sum(\cdot)$) and an activation function ($\sigma(\cdot)$) to get the updated new features ($h_i^{(l+1)}$), where $h_i^{(1)} = f_{i}$ in the input layer.
A $n$-layer network allows for message passing across $n$-hop in the graph. 
Therefore, the representation of a node$_x$ learned by a $n$-layer \texttt{BRGCN} model aggregates node features from a $n$-hop subgraph centered on node$_x$.
Besides, the different sets of weights $W_{r}^{d(l)}$ for different types of edges and sum-aggregation adopted in Equation~\ref{eq:rgcn2} can help to learn the graph structures corresponding to information flow and variable adjacency, respectively.
By learning different graph structures and node features, we believe that the network can distinguish vulnerable and benign nodes.

To train the model, we minimize the following cross-entropy loss on all labeled nodes:
\begin{equation}
\label{eq:loss}
\min_\theta\mathcal{L}=- \sum_{G \in \mathcal{G}} \frac{1}{|\mathcal{Y}|} \sum_{i \in \mathcal{Y}} \sum_{k=1}^{K} w_k \cdot y_{i k} \ln h_{i k}^{(L)},
\end{equation}
where $G$ is a graph in the training set $\mathcal{G}$, $\mathcal{Y}$ is the set of nodes in our training samples. $h_i^{(L)}$ is the output of BRGCN for node $i$. Note that we used softmax function for the last layer. Thus $h_i^{(L)}$ denotes the predicted class distribution for node $i$ with $h_{i k}^{(L)}$ being the probability of node $i$ belonging to class $k$, $k = \{0,1\}$.
$w_k$ is the weight for class $k$ and $y_{i k}$ denotes respective ground truth label for node $i$. 
We introduce $w_k$ in our loss function because class distribution in \texttt{DFG+} is extremely imbalanced, i.e., the majority of the nodes are negative nodes (benign nodes), while the positive nodes (vulnerable nodes) only take up a very small portion. To avoid the majority class dominate the loss function, we assign larger weight to positive class. 
$\theta = \{W_0^{(l)}, W_r^{in(l)}, W_r^{in(l)}; r \in \mathcal{R}\}_{l=1}^L$ is the set of the model parameters. After the loss is calculated in each training epoch, backward propagation computes gradient of the loss function with respect to the trainable parameters $\theta$, then parameters are updated to minimize loss.


\mypara{Model Parameter Size and Time Complexity}
For simplicity of the analysis, we first define the dimensionality of $W_r^{in(l)}$ and $W_r^{out(l)}$ as $W_r^{in(l)} \in \mathbb{R}^{d_l \times d_{l+1}}$ and $W_r^{out(l)} \in \mathbb{R}^{d_l \times d_{l+1}}$, where $d_l$ and $d_{l+1}$ is the dimensionality of the node representation in the $l$-th layer and $(l+1)$-th layer, respectively. Since $\theta = \{W_0^{(l)}, W_r^{in(l)}, W_r^{in(l)}; r \in \mathcal{R}\}_{l=1}^L$ is the set of parameters for BRGCN, the model parameter size is $\mathcal{O}(\sum_{l=1}^{L} \sum_{r \in \mathcal{R}} d_l \cdot d_{l+1})=\mathcal{O}(\sum_{l=1}^{L}  d_l \cdot d_{l+1} \cdot |\mathcal{R}|)$.

For the forward pass of \texttt{BRGCN}, the main time complexity in the $l$-th layer for node $v_i$ is the calculation of Equation~\ref{eq:rgcn2}, which is $\mathcal{O}( \sum_{r \in \mathcal{R}} (|\mathcal{IN}_i^{r}|+|\mathcal{OUT}_i^r|) \cdot d_l \cdot d_{l+1})$. It is equivalent to $\mathcal{O}(D_i \cdot d_l \cdot d_{l+1})$, where $D_i = \sum_{r \in \mathcal{R}} (|\mathcal{IN}_i^{r}|+|\mathcal{OUT}_i^r|)$ is the summation of the in-degree and out-degree of node $v_i$. Thus, the time complexity of BRGCN for node $v_i$ is $\mathcal{O}(D_i \cdot \sum_{l=1}^L  d_l \cdot d_{l+1})$. Then, the computational cost of \texttt{BRGCN} for a \texttt{DFG+} graph is $\mathcal{O}(\sum_{i} D_i \cdot \sum_{l=1}^L  d_l \cdot d_{l+1})$, which  is equivalent to $\mathcal{O}(|\mathcal{E}| \sum_{l=1}^L  d_l \cdot d_{l+1})$, where $\mathcal{E}$ is the set of edge in \texttt{DFG+}. The complexity of the backward propagation via gradient descent is the same as the forward pass. Thus, the total cost in one iteration is $\mathcal{O}(|\mathcal{E}| \sum_{l=1}^L  d_l \cdot d_{l+1})$.
\subsection{Vulnerability Identification}
\label{sec:iden}
In this subsection, we demonstrate how the trained model achieve the goal as proposed in Section~\ref{sec:prob}.
In the training phase, we leverage source code of vulnerable programs, inputs that can trigger the vulnerabilities, and the tools that we present in early subsection to generate labeled \texttt{DFG+} and save the corresponding supporting data.
Then we train the model with the generated \texttt{DFG+} from the vulnerable execution.
Through the forward prorogation (message passing) rules defined in Equation~\ref{eq:rgcn2}, loss function defined in Equation~\ref{eq:loss}, and backward prorogation to update the trainable parameters in the model, we get an effective model which can predict labels for nodes an given unlabeled \texttt{DFG+}.


In the testing phase we apply the trained model to predict the labels for unlabeled \texttt{DFG+} generated from binary-only software. Through the maps in the supporting data created by the runtime analyzer, the vulnerable nodes in \texttt{DFG+} can be mapped to corresponding instruction addresses in binary code and execution trace. Since the execution trace contains the address of memory operands, the address of corrupted variable can also be identified.
Note that in some cases that one execution may trigger several silent BOF vulnerabilities, the vulnerable instructions and corrupted variables can be identified separately.







\vspace*{-2mm} 
\subsection{Implementation}
We implement our system on 32-bit Linux system with Intel x86 Instruction Set Architecture.
The compiler plugin is built based on the LLVM-5.0.0 and its runtime library is built on runtime-rt-5.0.0.
The plugin and runtime library consists of 82 lines of new code and 2236 lines of new code, respectively, when comparing with implementation of ASAN.
The dynamic binary analyzer and graph constructor are developed on the Intel Pin 3.10., which consists of 9900 lines of C++ code in total.
The graph model consists of 800 lines of Python code, and is implemented base on the DGL v0.4.3~\cite{wang2019dgl}, an high performance and salable Python package for deep learning on graph typed data.

\vspace*{-2mm} 
\section{Experiment and Results}
\label{sec:exp}
\subsection{Data Generation and Preprocessing}
We select 30 reproducible CVEs as shown in Table~\ref{tab:cves} from a repository of Linux vulnerabilities~\cite{cves}. 
We generate three labeled \texttt{DFG+} for each CVE, from three different executions:
In the first execution, we compile the program by the compiler plugin and find an input to trigger the vulnerability.
In the second execution, we change the input which overflow the buffer with different length. 
In the third execution, we change the length of vulnerable buffer in the program's source code, recompile it to binary through our compiler plugin, and run the modified program again.
In all three executions, inputs are able to trigger the vulnerable without crashing the execution.
Since the length of vulnerable buffer or input can be hardly be changed in some programs,
we finally get 86 labeled \texttt{DFG+}s with over 35 millions (35084810) nodes, of which only 6708 nodes are positive.

We observe that the constructed \texttt{DFG+} vary largely in size (number of nodes), 
from a few thousands to a few millions.
It is impossible to fit an entire \texttt{DFG+} into \texttt{BRGCN} 
for end-to-end training, especially for those \texttt{DFG+} with more than 3 millions of nodes.
To alleviate this problem, we propose a graph cutting algorithm (the detaill of the Algorithm~\ref{alg:cut} is in the appendices). 
In the cutting algorithm, we firstly cut a big graph into several small graph by removing edges that connect different sub-graphs,
all the nodes in the sub-graphs is \textit{sample nodes}.
Secondly, we add $n$-hop neighbors to each sub-graph as {\em supporting node}, where $n$ is the number of model layers.
When training model on the sub-graphs, both \textit{supporting node} and \textit{sample nodes} are involved in
forward propagation whereas only the \textit{sample nodes} was considered for calculating loss.

\begin{table}[h]
    \centering
    \footnotesize
    \caption{Information and testing results of each CVE cases.}
    \label{tab:cves}
    \begin{tabular}{l|cc|c}
    \toprule[0.5pt]
    \toprule[0.5pt]
    
    \multicolumn{3}{c|}{\bf{\emph{Vulnerability Information}}} & \multicolumn{1}{c}{\bf{\emph{Analysis Result}}}
     \\ \hline
     {\tt CVE-ID} & \parbox[t]{0.5cm}{\centering {\tt Name}} & \parbox[t]{1cm}{\centering {\tt Region}} & \parbox[t]{0.8cm}{\centering {\tt Detected}} 
     \\ \hline
    \rowcolor{tablegray}
    CVE-2004-0597	& pngslap   & stack & \ding{51} \\
    
    CVE-2004-1120 & proz    & stack	&   \ding{51}\\
    
    \rowcolor{tablegray}
    CVE-2004-1255	& 2fax	& stack &   \ding{51}\\
    
    CVE-2004-1257	& abc2mtex	& stack &   \ding{51}\\
    
    \rowcolor{tablegray}
    CVE-2004-1261 & asp2php	& stack	&  \ding{51}\\
    
    CVE-2004-1262 & bsb2ppm	& stack	& \ding{51} \\
    
    \rowcolor{tablegray}
    CVE-2004-1275	 & html2hdml  & stack	&  \ding{51}  \\
    
    CVE-2004-1278	& jcabc2ps	& stack	&  \ding{51}\\
    
    \rowcolor{tablegray}
    CVE-2004-1279    & jpegtoavi  & stack & \ding{51} \\
    
    CVE-2004-1287	 & nasm	& stack & \ding{51}	\\
    
    \rowcolor{tablegray}
    CVE-2004-1288	 & o3read	   & stack &  \ding{51}\\
    
    CVE-2004-1289	& pcal	& stack 	&  \ding{51}\\
    
    \rowcolor{tablegray}
    CVE-2004-1290	& pgn2web    & stack  	&  \ding{51}\\
    
    CVE-2004-1292	& ringtonetools	& stack 	& \ding{51} \\
    
    \rowcolor{tablegray}
    CVE-2004-1293	& rtf2latex2e.bin & stack	&  \ding{51}\\
    
    CVE-2004-1297   & unrtf	& stack  &  \ding{51}\\
    
    \rowcolor{tablegray}
    
    CVE-2004-2093	& rsync  & stack  & \ding{51} \\
    CVE-2004-2167 & latex2rtf	   & stack  & \ding{55}  \\
    \rowcolor{tablegray}

    CVE-2005-0101   & newspost	& stack  	& \ding{51} \\
    CVE-2005-3862	 & unalz	   & stack   	& 	\ding{51}\\
    \rowcolor{tablegray}

    CVE-2005-4807	& as-new & stack     & \ding{51}  \\
    CVE-2007-1465 & dproxy	   & stack  & \ding{51} 	\\
    \rowcolor{tablegray}

    CVE-2009-1759   & ctorrent	& stack  	&  \ding{51}\\
    CVE-2009-2286	 & compface	   & stack 	& \ding{51}	\\
    \rowcolor{tablegray}

    CVE-2009-5018	& gif2png & stack     & \ding{51}  \\
    CVE-2010-2891   & smisubtree	& stack  	&  \ding{51}\\

    \rowcolor{tablegray}

    EDB-890 & psnup	   & stack	& \ding{51}\\
    EDB-9264 & stftp	   & stack	& \ding{51}\\
    
    \rowcolor{tablegray}

    EDB-14904   & fcrackzip	& stack  	&  \ding{51}\\
    EDB-15062 & rarcrack	& stack	    & \ding{51}\\
    \bottomrule[0.5pt]
    \bottomrule[0.5pt]
    \end{tabular}
    
    \vspace{-2ex}
\end{table}
As can be noticed above, the dataset is extremely imbalanced~\textendash~the ratio 
between positive and negative nodes is roughly 1/5230.
To further reduce the number of negative nodes, 
we exclude all the \texttt{r-node} and \texttt{i-node} in \textit{sample nodes}
because BOF can only overwrite variables in memory. 
We also exclude nodes without any incoming \texttt{d-edge} because the \textit{live} variables associated with 
vulnerable nodes must be written through \textit{invalid operation} in BOF.
After the exclusion, we are able to reduce the ratio to 1/659.
Note that by excluding we mean we won't choose them as \textit{sample nodes} in sub-graphs.
Instead, we select them as \textit{supporting nodes} if they are the neighbors of \textit{sample nodes} 
to help classify the \textit{sample nodes} in sub-graphs.
Finally, we further reduce the number of negative through random sample.

\subsection{Evaluation}
We experimented with different number of layers, size of hidden states, and dropout rates to find the best performing model. 
Currently \texttt{BRGCN} has 4 layers including an input layer and an output layer and each layer has hidden states with dimension 16. 
10 sets of parameters ($W$) are used for 5 types for edges (2 sets of parameters for each type).

After we get the best configurations, we adopt 8-fold cross-validation to comprehensively evaluate the model.
In each round of the cross-validation, we select 75\%, 12.5\% and 12.5\% of 86 graphs as training set, 
validation set and testing set.
Table~\ref{tab:brgcn} presents the \textit{Accuracy}, \textit{Precision}, \textit{Recall} and \textit{F1} on the test set.
Our model achieve 94.39\% accuracy and 94.18\% F1 score on the sampled dataset.
Since we are the first one to analyze the silent vulnerability through deep learning, 
we cannot find the similar works to compare. 
However, we will compare our design with some other potential designs in next section.

\begin{table}[b]
    \centering
    \footnotesize
    \caption{The overall performance of our proposed models.}
    \label{tab:brgcn}
    \begin{tabular}{c|c|c|c|c}
    \hline
    \bf{\emph{Fold}} & \bf{\emph{Accuracy}} & \bf{\emph{Precision}}  & \bf{\emph{Recall}} & \bf{\emph{F1}} \\
    \hline
    fold-1 & 0.8871 & 0.9828 & 0.7703 & 0.8636  \\
    \hline
    fold-2 & 0.9623 & 1.0000 & 0.9298 & 0.9636 \\
    \hline
    fold-3 & 0.9712 & 0.9455 & 1.0000 & 0.9719  \\
    \hline
    fold-4 & 0.9072 & 0.9167 & 0.8958 & 0.9060  \\
    \hline
    fold-5 & 0.9617 & 0.9657 & 0.9574 & 0.9615  \\
    \hline
    fold-6 & 0.9503 & 0.9244 & 0.9821 & 0.9523  \\
    \hline
    fold-7 & 0.9359 & 0.8864 & 1.0000 & 0.9397  \\
    \hline
    fold-8 & 0.9757 & 0.9537 & 1.0000 & 0.9763 \\ 
    \hline
    Average& 0.9439 & 0.9469 & 0.9419 & 0.9418 \\
    \hline
  \end{tabular}
  \end{table}

Then, we examine our model's ability to identify vulnerable operations in silent BOFs.
Since a silent BOF will result in one or more vulnerable nodes, 
we can successfully locate the vulnerable operation as long as one vulnerable node is identified. 
As a result the vulnerabilities detection rate is much better than the vulnerable node detection rate.
Table~\ref{tab:cves} shows the detection results when we map the vulnerable node in test phase to the executables. 
Due the limited numbers of global/heap buffer overflow in vulnerability database, 
we did not find a reproducible one in our evaluation.
But we modify vulnerable stack buffers in several cases displaied in Table~\ref{tab:cves} to global/heap buffers, 
ensure the buffer overflow corrupted some adjacent variables, 
and successfully identify them through our trained model.
In summary, our model successfully identify and locate 29 vulnerabilities out of 30 CVE traces.
The evaluation result indicate that the vulnerable patterns in data flow level for stack BOF, global BOF and heap BOF are similar.

\vspace*{-2mm} 
\section{Explainability}
\label{sec:explain}
When designing \texttt{DFG+} and \texttt{BRGCN}, we raise several insights based on our intuitions, in this section we try to explain their effectiveness through several experiments.
Accordingly, we put forward several evaluation questions:
1) Can sequenced model such as RNN and LSTM solve the problem through analysis on the instruction sequence directly? 
2) If a homogeneous graph, rather than a more complex relational graph, is enough to classify vulnerable nodes in \texttt{DFG+}?
3) Is \texttt{BRGCN} defined in Equation \ref{eq:rgcn2} more effective than RGCN defined in Equation~\ref{eq:rgcn}?
4) Can \texttt{BRGCN} effectively identify vulnerable nodes in traditional data flow graphs?
5) Can \texttt{BRGCN} effectively identify vulnerable nodes in \texttt{DFG+} with ID as node attributes?
6) Does \texttt{BRGCN} really benefit from training on multiple graphs?

To answer these questions, we setup 4 groups of experiments. 
In the first group of experiments, we firstly generate instruction trace 
which includes executed instructions and access memory addresses, 
secondly split the instruction trace into fix-length sequence to make them end with memory access instructions.
Thirdly, if the last instruction of a sequence results in vulnerable operation in silent BOF, 
we label this sequence as vulnerable sequence, otherwise we label it as benign sequence.
Fourthly, we sample the same positive samples and negative samples as that sampled in training \texttt{BRGCN}. 
Finally, we adopt an open source implementation of Memory-Augmented RNN and LSTM~\cite{memoryrnn} 
to classify execution traces and the results are reported in Table~\ref{tab:explain}.
From the experiment results, we can easily conclude that RNN and LSTM 
cannot help to identify vulnerable operation by analyzing instruction sequence with access memory addresses.

In the second group of experiments, we adopt ConvGNN and RGCN and train models on \texttt{DFG+}.
In the ConvGNN, all types of edges are treated homogeneously and processed with the same weight matrix $W$. 
The RGCN adopts different propagation rules for different edge types, and propagate node features along with the incoming direction of edges.
The experiment results in Table~\ref{tab:explain} shows that the performance of \texttt{BRGCN} is better than RGCN, and the performance of RGCN is better than ConvGNN. 
This indicates that adopting two sets of parameters for each type of edge is more effective than one set of parameter for each type and a single set of parameter regardless of edge types.

In the third group of experiments, we change the structure of \texttt{DFG+}.
There are two variants: 1) graphs with only program runtime data flow, and 2) graphs with nodes that are assigned unique IDs as node attributes.
Then, we train \texttt{BRGCN} model on the two sets of modified graphs and display
their result in Table~\ref{tab:explain}.
From the results we know:
1) the \texttt{BRGCN} cannot distinguish significant difference between local graph structures of \textit{invalid operations} and other benign operations in data flow graph only, 
the adoption of spatial information and other implicit information flow plays an important role for the node classification problem
and
2) when training a model on different graphs, the adopting of node IDs as node attributes is harmful.

In the last group of experiments, we try to train our model on graphs generated from a single program and test it on other programs.
The evaluation result shows that the model trained on a single program is significantly worse than model trained on multiple programs. 
We conclude that by carefully designing the \texttt{DFG+}, our model can benefit from different programs/graphs.
It indicates common semantic features for BOF vulnerabilities are shared by different programs.  

\begin{table}[t]
    \centering
    \captionsetup{justification=centering}
    \footnotesize
    \caption{The performance comparison of different neural networks and graph structures.}
    \label{tab:explain}
    \makeatletter
    \newcommand{\widenhline}{%
        \noalign {\ifnum 0=`}\fi \hrule height 0.5pt
        \futurelet \reserved@a \@xhline
    }
    \newcolumntype{I}{!{\vrule width 1pt}}
    \makeatother
    \begin{tabular}{m{0.7cm}<{\centering}Im{1.7cm}<{\centering}|c|c|c|c}
    \hline
     Group &\bf{\emph{Setting}} &  \bf{\emph{Accuracy}} & \bf{\emph{Precision}}  & \bf{\emph{Recall}} & \bf{\emph{F1}} \\
    \hline
    \multirow{2}{*}{\emph{1}}  & RNN  & 0.4977 & 0.4994 & 0.5557 & 0.5260 \\
    \cline{2-6} 
    & LSTM  & 0.4948 & 0.4929 & 0.5136 & 0.5030 \\
    \hline
    \multirow{2}{*}{\emph{2}} & ConvGCN & 0.7914 & 0.8105 & 0.7619 & 0.7616    \\
    \cline{2-6} 
    & RGCN & 0.8411 & 0.8699 & 0.8158 & 0.8175  \\
    \hline
    \multirow{3}{*}{\emph{3}} &\texttt{BRGCN} w/ DF-Only & 0.7001 & 0.6126 & 0.7702 & 0.6793  \\
    \cline{2-6} 
    & \texttt{BRGCN} w/ Node-ID & 0.7686 & 0.7769 & 0.7466 & 0.7577  \\
    \hline
    \multirow{1}{*}{\emph{4}} & \texttt{BRGCN} w/ One-Program & 0.5741 & 0.4839 & 0.3562 & 0.4215\\
    \hline
  \end{tabular}
  \vspace{-2ex}
  \end{table}

\section{Limitations and Conclusion}
\label{sec:con}
Our approach suffers from several limitations. First, although we achieve high detection rate of silent BOFs, there is considerable false-positive predictions on nodes, meaning that some benign nodes are falsely classified as vulnerable nodes. This is inherent from the extreme imbalance of the number of positive and negative nodes, which has a ratio as low as $0.0002$. We have tried to down-sample the negative nodes and apply a class-weighted loss function during training, it is still an issue because any single percentage drop of classification precision would lead to considerable false positive predictions. Second, although our model can test a graph very quickly (less than 0.2 seconds on average), there is an overhead in collecting program runtime data flow and building \texttt{DFG+}. Unless we can trace program data flow on the fly, it is not practical to deploy our framework for detecting vulnerabilities in real time production environment.

In this paper, we design a novel graph data structure \texttt{DFG+} to represent the program's runtime information flow and variables' spatial information. 
A runtime analyzer is implemented to construct \texttt{DFG+} and the AddressSanitizer is customized to help label the nodes.
We further propose \texttt{BRGCN} to analyze \texttt{DFG+} and detect vulnerable nodes with 94.39\% accuracy. 
Through mapping of the vulnerable nodes back to the execution trace, we are able to locate the vulnerable points in the program at the binary level. 
We believe the \texttt{DFG+} and the \texttt{BRGCN} proposed in our work have wide applications.
Our proposed scheme could be used in vulnerability analysis to help locate vulnerable point, in software patch to help generate patches in the binary, in exploit generation to help attack vulnerable software, and in software testing to help find software bugs.

Finally, we would like to suggest some future works that could supplement our approach. Some possible avenues include applying the GNN-based approach to other silent vulnerable executions such as detecting buffer overread or on obfuscated programs.

\bibliographystyle{IEEEtran}
\bibliography{main}
\newpage
\begin{appendices}
\label{sec:apdx}

\section{Graph Cut Algorithm}

\newcommand\stackequal[2]{%
  \mathrel{\stackunder[2pt]{\stackon[4pt]{=}{$\scriptscriptstyle#1$}}{%
  $\scriptscriptstyle#2$}}}
\begin{algorithm}[!htb]
    \textsc{Input}: graph $\mathbf{G}$($\mathcal{N}$, $\mathcal{E}$); number of layer $l$; number of subgraph $n$; \\
    \textsc{Output}: a set with $m$ subgraph: $\mathbf{C} = \left\{G_i \mid 0 \leqslant i < n \right\}$, 
    and the IDs of sampled nodes $S_i$ in each subgraph $G_i$; 
 
    \begin{algorithmic}[1]
    \caption{Graph Cut Algorithm}\label{alg:cut}
        \STATE initialize a set of $n$ subgraph: $\mathcal{C} = \left\{G_i \mid 0 \leqslant i < n \right\}$, 
        where $G_i$ = ($N_i$, $E_i$), $N_i$ = $\varnothing$, $E_i$ = $\varnothing$;
        \STATE divided nodes in $\mathbf{G}$ into $m$ samples: $S_i \mid 0 \leqslant i < n $, satisfying $\left| S_i \right| \leqslant \lceil \left| \mathcal{N} \right|/m \rceil$ 
        \AND
        $\mathrel{\stackunder[2pt]{\stackon{$\cup$}{\scriptsize n-1}}{\scriptsize i=0}}S_i = \mathcal{N} $  
        \FOR {$i=n$ down to 1}
        \STATE $N_i$ := $S_i$
        \FOR {$j=l$ down to 1}
        \FOR {$e \in \mathcal{E}$}
        \STATE /* $src(e)$, $des(e)$ denotes the source node and destination node of $e$ */
        \IF{$src(e) \in N_i$ \AND $des(e) \notin N_i$}
        \STATE add $dst(e)$ to $N_i$;
        \ENDIF
        \IF{$src(e) \notin N_i$ \AND $des(e) \in N_i$}
        \STATE add $src(e)$ to $N_i$
        \ENDIF
        \ENDFOR
        \ENDFOR
        \FOR {$e \in \mathcal{E}$}
        \IF{$src(e) \in N_i$ \OR $dst(e) \in N_i$}
        \STATE add $e$ to $E_i$;
        \ENDIF
        \ENDFOR
        \ENDFOR
      \end{algorithmic}
    \end{algorithm}

\end{appendices}

\end{document}